\documentclass[aps,12pt]{revtex4-1}
\usepackage{amsmath}
\usepackage{graphicx}

\begin{document}

\title{Reanalysis of experimental determinations of polariton-polariton interactions in microcavities}

\author{D. W. Snoke$^1\footnote{Email: snoke@pitt.edu}$, V. Hartwell$^2$, J. Beaumariage$^1$,  S. Mukherjee$^1$, Y. Yoon,$^3$ D. M. Myers$^4$,  M. Steger$^5$,  Z. Sun,$^1$ 
K. A. Nelson$^6$, and L. N. Pfeiffer$^7$ }

\affiliation{$^1$Department of Physics and Astronomy, University of Pittsburgh, 3941 O'Hara St., Pittsburgh, PA 15260, USA\\
$^2$Skibo Energy Systems, 6901 Lynn Way, Pittsburgh, PA 15208, USA\\
$^3$Berkeley and Lawrence Berkeley National Laboratory, University of California at Berkeley, Berkeley, California 94720, USA\\
 $^4$Department of Engineering Physics, Polytechnique Montr\'eal, Montr\'eal, QC H3C 3A7, Canada\\
 $^5$National Renewable Energy Lab, 15013 Denver Parkway, Golden, CO 80401, USA\\
 $^6$Department of Chemistry, Massachusetts Institute of Technology, 77 Massachusetts Avenue, Cambridge, MA 02139, USA\\
 $^7$Department of Electrical Engineering, Princeton University, Princeton, NJ 08544, USA
}

\begin{abstract}
The polariton-polariton interaction strength is an important parameter for all kinds of applications using the nonlinear properties of polaritons, such as optical switching and single-photon blockade devices.  In this paper, we review and compare the results of a series of experiments on polariton-polariton interactions in GaAs/Al$_x$Ga$_{1-x}$As microcavity polariton structures, and present new theoretical analysis of these experiments.  We show that not just the energy shift of the spectral lines, but also the results of measurements sensitive to the polariton scattering rate are important for the calibration of the interaction parameter at low excitation density.  We find that when adjustments are made to correct for new understanding of the experiments, the value of the interaction parameter at low density is lower than previous reported, but still significantly higher than theoretically predicted. 

\end{abstract}

\maketitle

\section{Introduction}

The field of optics of microcavity polaritons has exploded in recent years, driven both by the fundamental interest in Bose-Einstein condensation (BEC) of polaritons (for reviews of previous and current work on BEC of polaritons, see Refs.~\onlinecite{littlewood} and \onlinecite{keeling} and references therein), and also the highly nonlinear properties of these systems, which may allow, for example, single-photon blockade effects \cite{Delteil2019,Matutano2019} and novel optoelectronic devices \cite{baum,ballarini,recentgate}. Both the BEC effects and the nonlinear effects for applications rely crucially on the magnitude of the polariton-polariton interaction strength. 

In terms of the Gross-Pitaevskii equation for condensates, this interaction strength is parametrized as the constant $g$,
\begin{equation}
i\hbar\frac{\partial \psi}{\partial t} = 
-\frac{\hbar^2}{2m} \nabla_{\|}^2\psi + g|\psi|^2\psi,
\label{GP1}
\end{equation}
while in terms of optics, the same equation becomes the nonlinear wave equation (for the derivation, see, e.g., Ref.~\onlinecite{snokebook2}, Section 11.13),
\begin{equation}
i\hbar\frac{\partial \psi}{\partial t} = 
-\frac{\hbar^2}{2m} \nabla_{\|}^2\psi- \frac{2\mu_0\chi^{(3)}(\hbar\omega)^2}{m}|\psi|^2\psi,
\end{equation}
where the $\chi^{(3)}$ term determines the strength of the effective particle-particle interaction.  The form of the Gross-Pitaevskii equation (\ref{GP1}) corresponds in many-body theory to the interaction Hamiltonian, defined for a two-dimensional polariton system,
\begin{equation}
H_{\rm int} =  \frac{1}{2S} \sum_{{\vec p},{\vec q},{\vec k}} U
a^{\dagger}_{{\vec p}} a^{\dagger}_{{\vec q}} a^{ }_{{\vec q}+ {\vec k}} a^{ }_{{\vec p}
- {\vec k}},
\label{bogoh}
\end{equation}
where $U$ is the interaction strength, $a^{\dagger}_{\vec{k}}$ and $a^{ }_{\vec{k}}$ are the field creation and destruction operators, respectively, and $S$ is the area. Second-order perturbation theory then gives the renormalized particle energy due to interactions as (cf.~Ref.~\onlinecite{snokebook2}, Section 8.1),
\begin{eqnarray}
\Sigma_i = \langle i | H_{\rm int} | i\rangle + \sum_{n \ne i} \frac{|\langle n | H_{\rm int}| i\rangle|^2}{E_i-E_n} 
+ \frac{i\pi }{\hbar} \sum_{n \ne i} |\langle n | H_{\rm int}| i\rangle|^2 \delta(E_i-E_n).
\label{intE}
\end{eqnarray}
The first term on the right-hand side is known as the {\em mean-field} energy shift due to interactions, and is normally presumed to be much larger in magnitude than the real part of the second term, which is negative. For a condensate, the mean-field energy is simply $Un$, where $n$ is the particle density, while for non-condensed particles, exchange energy adds another, equal term, giving $\Sigma E = 2Un$, which is typically written $E = gn$. The third, imaginary term on the right-hand side gives Lorentzian line broadening, proportional to the particle-particle scattering rate (for a proof see Section 8.4 of Ref.~\onlinecite{snokebook2}). Exchange for bosons also occurs in this factor for a non-condensed gas, so that the scattering rate is proportional to $(2U)^2$ (cf.~Ref.~\onlinecite{snokebook2}, Section 4.8); i.e., the same factor $g$ is used. 

The above analysis implies several different experimental methods by which the interaction strength $g$ can be estimated. One is to measure the shift of the ground state energy as a function of density. Another is to measure the line broadening of the particle energy as a function of density. An advantage of optical polariton systems is that both of these are directly observable in the spectroscopic data, although there are complicating factors, as discussed below. Another method of finding $g$ is to deduce the particle-particle scattering rate from nonequilibrium particle distributions, as discussed in Section IV. Finally, the interaction strength can also be estimated from the temporal correlation function of the polaritons, as discussed in Section V. 

\begin{figure}
\centering
\includegraphics[width = .5\linewidth]{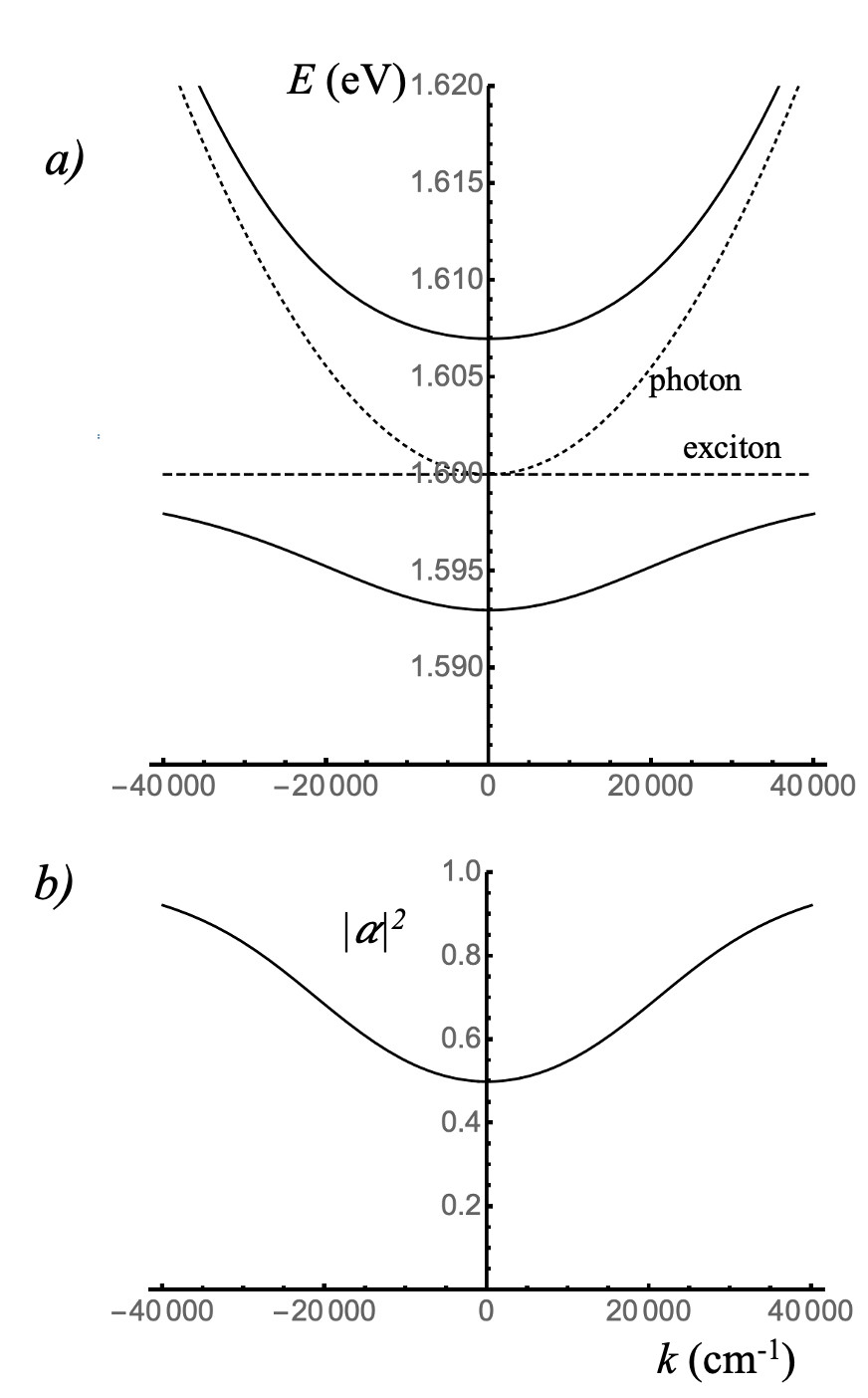}
\caption{Typical polariton properties for a GaAs microcavity structure like those used in the experiments discussed here. a) Dashed line: bare exciton energy. Dotted line: bare cavity photon energy, given by $E_c = \hbar c \sqrt{k_0^2 + k^2}$. Solid lines: the upper and lower polariton branches formed by mixing the exciton and photon states with a Rabi coupling of $\Omega = 7$ meV and index of refraction $n = 3$. b) The exciton fraction of the lower polariton branch corresponding to the dispersion of (a). } 
\label{mc1}
\end{figure}

The basic theory of polariton-polariton interactions assumes that they interact entirely through their excitonic component. One writes the polariton state as
\begin{equation}
|{\rm pol}\rangle = \alpha|{\rm ex}\rangle + \beta|{\rm phot}\rangle,
\label{polfrac}
\end{equation}
where $|{\rm ex}\rangle$ and $|{\rm phot}\rangle$ are pure exciton and photon states, respectively, and $\alpha$ and $\beta$ are complex factors that depend on the polariton momentum and the energy difference between the photon and exciton (the ``detuning'').  Figure \ref{mc1} shows the energy and exciton fraction $|\alpha|^2$ of the polariton states as a function of $k$, for a typical microcavity structure when the bare exciton and cavity photon energies are tune to be equal at $k=0$. As seen in this figure, the lower polariton states evolve continuously into exciton states at higher energy and momentum. Therefore, when the particles scatter into different states, they also change their character of how excitonic they are.  If we write the pure exciton-exciton interaction as $g_{\rm ex}$, then because the interaction (\ref{bogoh}) has four polariton operators, the interaction strength between polaritons will be $g = |\alpha|^4g_{\rm ex}$. (In general, $\alpha$ is a function of $k$, which means that there can be four different values of $\alpha$ involved in a given two-body collision, but we can generally take an average value of $\alpha$ based on the detuning of the polaritons near $k=0$.)

When multiple, degenerate quantum wells are embedded in a single microcavity, a single polariton will be a superposition of one photon and an exciton in each quantum well.  The coefficient of any single quantum-well exciton state within the polariton state will be $\alpha/\sqrt{N_{QW}}$, where $N_{QW}$ is the number of quantum wells, and $\alpha$ is the excitonic coefficient for a single quantum well polariton with the same exciton-photon detuning.  If we assume that excitons in separate quantum wells do not interact, then for a multiple quantum well system we write $g = |\alpha|^4g_{\rm ex}/N_{QW}$. We can also define an effective polariton-{\em exciton} scattering rate $\tilde g = |\alpha|^2 g_{\rm ex}/N_{QW}$ (i.e., $|\alpha|$ for the exciton states can be assumed $\sim 1$.)

The pure exciton-exciton interaction strength $g_{ex}$ has been calculated theoretically using a variational approach based on the exciton wave function in the Wannier limit \cite{tassone,ciuti,ciuti2,wouters}. The full calculation is difficult, and most prior work has made significant approximations such as bosonization (treating the excitons as pure bosons, which has been shown to ignore important exchange processes \cite{combescot}),  sticking only to a mean-field calculation (which ignores the possibilty of deformation of the excitons away from their ground-state 1$s$ orbital wave function), and ignoring electron-hole exchange altogether.  In general, one expects from unit analysis that the interaction constant should be independent of $k$ as $k\rightarrow 0$ and of the order of ${\rm Ry}_{\rm ex}a_B^2$, where ${\rm Ry}_{\rm ex}$ is the exciton Rydberg energy and $a_B$ is the exciton Bohr radius. The excitonic Rydberg is of the order of 10~meV in III-V semiconductor quantum wells, while the excitonic Bohr radius is of the order of 100~\AA~in the same materials. Specifically for GaAs/AlGaAs structures, which the majority of microcavity polariton experiments have used, Refs.~\onlinecite{tassone,ciuti,ciuti2,wouters} give $g_{ex} \sim 12-15$~$\mu$eV-$\mu$m$^2$ for spin-aligned excitons. This number is positive, corresponding to repulsive interactions, giving a mean-field energy shift upward, that is, a ``blue shift,'' as density increases.  Polaritons with opposite spin have a much weaker interaction, which can be neglected to lowest order.  For a polariton gas with randomized spin, this will give an average interaction strength $\sim 6$~$\mu$eV-$\mu$m$^2$. On the other hand, as discussed above, boson exchange in a nondegenerate gas gives an extra factor of 2, so that the proper number to use for a nondegenerate gas is $g_{ex} \sim 12$~$\mu$eV-$\mu$m$^2$.

Assuming no interaction between different quantum wells, for a GaAs structure with 12 quantum wells and exciton fraction of 50\%, this implies a polariton-polariton interaction constant $g \sim 0.25~\mu$eV-$\mu$m$^2$.  For multiple-quantum well (MQW) structures with thin barriers between the wells, of the order of the exciton Bohr radius or less, the factor to divide by may be less than $N_{QW}$, giving a larger effective polariton-polariton interaction. In typical III-V structures of the type considered here \cite{yamamoto,science,bloch}, there are three groups of four quantum wells, with each group placed at an antinode of a $3\lambda/2$ cavity.  The width of the quantum wells is approximately 70~\AA~and the width of the barriers is approximately 30~\AA, compared to an exciton Bohr radius of the order of 100~\AA. This leads one to expect that excitons in adjacent quantum wells do interact, and therefore the dividing factor should not be the full 12 for the total number of quantum wells, but perhaps as low as 3. 


The experimental estimates of $g$ for polaritons in GaAs/Al$_x$Ga$_{-1x}$As microcavities, using different methods, have ranged over more than three orders of magnitude, many of which are one or two orders of magnitude larger than the theoretical estimate of $g \sim 0.25~\mu$eV-$\mu$m$^2$ discussed above (e.g., Refs. \onlinecite{vlad,ferrier,walker}). There are several complicating factors, the most important of which is the presence of ``bare'' excitons in states at higher energy, also known as the ``exciton reservoir.'' In {\em non-resonant excitation} experiments, a laser is tuned to a photon energy much higher than the polariton energy, creating hot carriers which then lose energy by phonon emission, landing in exciton states. These excitons can then scatter down into polariton states. Depending on the details of the experiment, there may be many more of these excitons than polaritons, as discussed below. The exact number of excitons is hard to measure directly, because many of the exciton states are non-light-emitting, or ``dark,'' either because of selection rules or because they have momentum outside the ``light cone.'' The light cone is given by the angle of incidence at which photons inside the cavity will be totally internally reflected. Because the angle of emission has a one-to-one mapping to the in-plane momentum of the polaritons, this means that there is a maximum momentum of the exciton-polaritons that can emit light. In addition to both of these effects, many experiments may simply not have collected light from those exciton states that do emit photons, by cutoffs in collection angle or photon energy. 

In {\em resonant excitation} experiments (e.g. Ref.~\cite{devres,devres2}), polaritons are created directly by tuning an external laser wavelength to match a polariton energy. One might naturally assume that no excitons will be generated in such an experiment, if the spectral width of the excitation laser is much less than the energy separation between the lower polaritons and the bare excitons, but it is also possible for polaritons to scatter up into exciton states that lie 5-7 meV higher in energy. Once they are there, they can have much longer lifetime than the polaritons, because they have much lower rate of photon emission. 

The presence of this background of excitons, or exciton ``reservoir,'' has been shown in numerous experiments (e.g., Refs. \onlinecite{baum} and \onlinecite{myers-pillar,ostroTF,ostrobogo,exdiff}). One evidence is the observation of a mean-field energy shift of the polariton line much greater than expected from the polariton density.  The exciton-polariton interaction strength is stronger than the polariton-polariton interaction strength, proportional to $|\alpha|^2$ instead of $|\alpha|^4$; if the exciton population is comparable to the polariton density, the energy shift of the polariton states at zero detuning due to the presence of excitons will be twice as large as the effect of polaritons on each other. 

\section{Spectroscopic measurements of $g$ at low density}

\begin{figure}
\centering
\includegraphics[width = .8\linewidth]{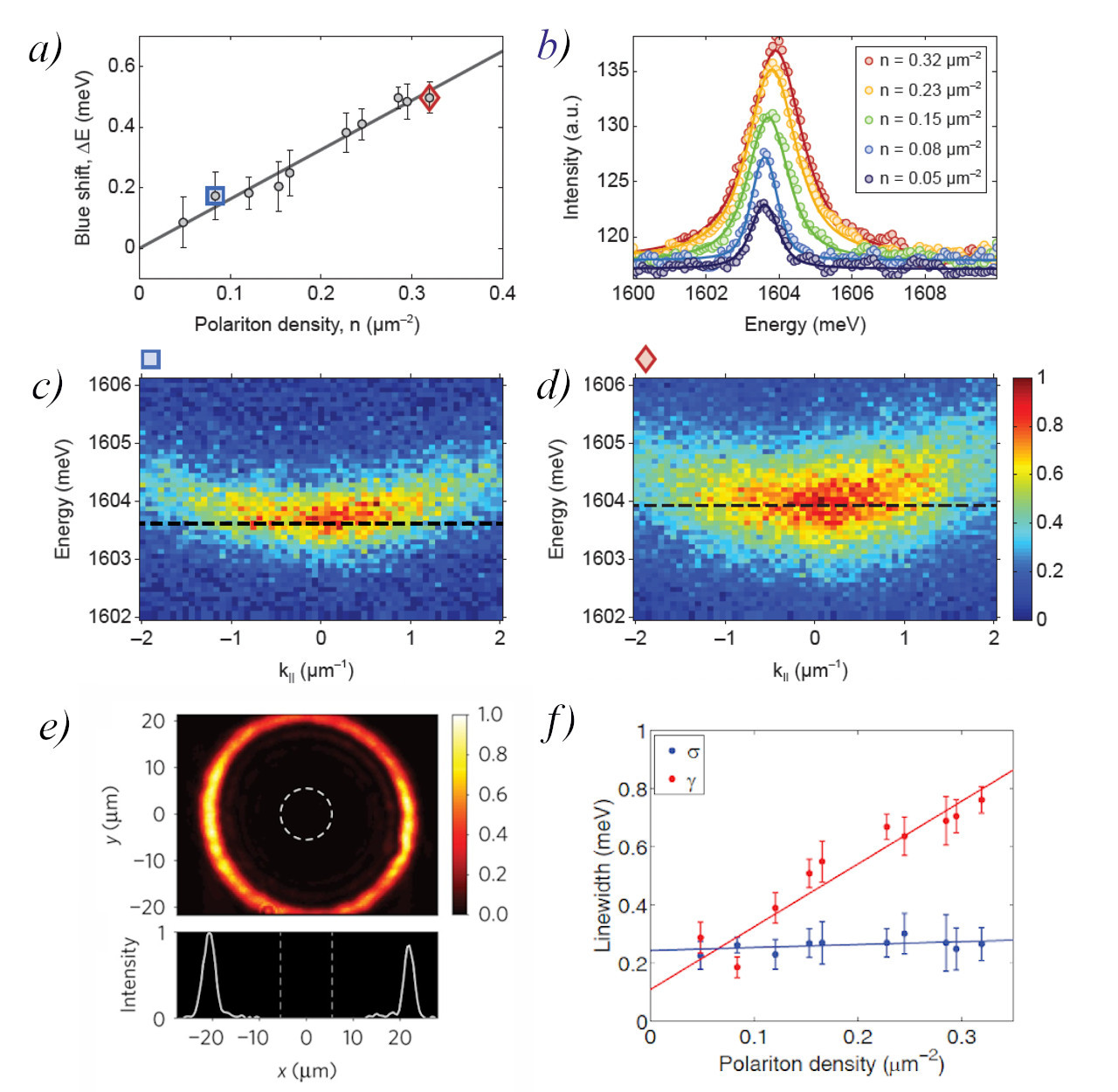}
\caption{a) Energy shift of the lower polariton line as a function of measured polariton density, taken from the peak position of spectra like those shown in (b). The images in (c) and (d) show typical momentum-resolved images from the data; the spectra of (b) correspond to vertical slices through these images at $k=0$.  e) Image of the excitation laser pattern generating the polaritons, and the laser intensity profile across a line through the center. f) Linewidth data from a Voigt fit to the same data, with a Gaussian part (blue symbols) convolved with a Lorentzian function (red symbols). From Ref.~\protect\onlinecite{naturephys}. } 
\label{mit2}
\end{figure}

Ref.~\onlinecite{naturephys} reported a study aimed at deducing the polariton-polariton interaction strength under conditions when the effect of the exciton reservoir was greatly reduced. Figure \ref{mit2}(e) shows an image of the circular laser pattern used to generate the polaritons in a GaAs-based microcavity. This pattern had two functions: first, the excitons created at the laser excitation ring formed a barrier that trapped polaritons in an equilibrium in the middle of the ring, without streaming away. Second, by putting the non-resonant exciton generation far from the center of the ring, the effect of excitons, which generally diffuse much more slowly than polaritons, was assumed to be negligible when observing only the polaritons at the center. 

Under the assumption of no excitons at all in the center of the ring, the results of Ref.~\onlinecite{naturephys} for the mean-field blue shift implied a surprisingly large value of the effective interaction constant. As seen in Figure \ref{mit2}(a), a substantial energy shift with density was seen. Great care was taken to measure the absolute density of the polaritons using a calibrated photon source. For the measured polariton densities, the reported interaction strength when extrapolated to the pure exciton limit corresponded to $g_{ex} \sim 1.7 $ meV-$\mu$m$^2$, many orders of magnitude larger than the theoretical prediction given above. 

Several needed corrections to these results were soon identified. One is the effect of quantum confinement on the kinetic energy of the polariton states. Even if there were no diffusion of the excitons at all, the barrier height of the ring would still increase as the exciton density increases. This in turn deters the tunneling of the polariton wave function into the barrier, effectively making the confined region smaller. When the wavelength of low-energy polaritons is comparable to the trap size, this can give a small but measurable blue shift of the polariton energy entirely due to the increase of the kinetic energy due to the quantum confinement. 

This effect had been seen before for very small traps with short-lifetime polaritons \cite{baum-shift}; the results of Ref.~\onlinecite{ferrier} also showed a strong blue shift that increase as the trap size decreased below 10~$\mu$m, indicating the effect of quantum confinement.
This effect was expected to be negligible for large traps or 40-50 $\mu$m diamter with long-lifetime polaritons, but Pieczarka and coworkers \cite{piec} showed that for polaritons with high photon fraction (high ratio of $|\beta/\alpha|$ in Equation (\ref{polfrac})), this effect can give a blue shift of up to about 50 $\mu$eV for the experiments of Ref.~\onlinecite{naturephys}. It cannot explain the much larger blue shifts at higher exciton fraction, however, nor the line broadening seen in Fig.~2(b), as discussed below. 

\begin{figure}
\includegraphics[width = .6\linewidth]{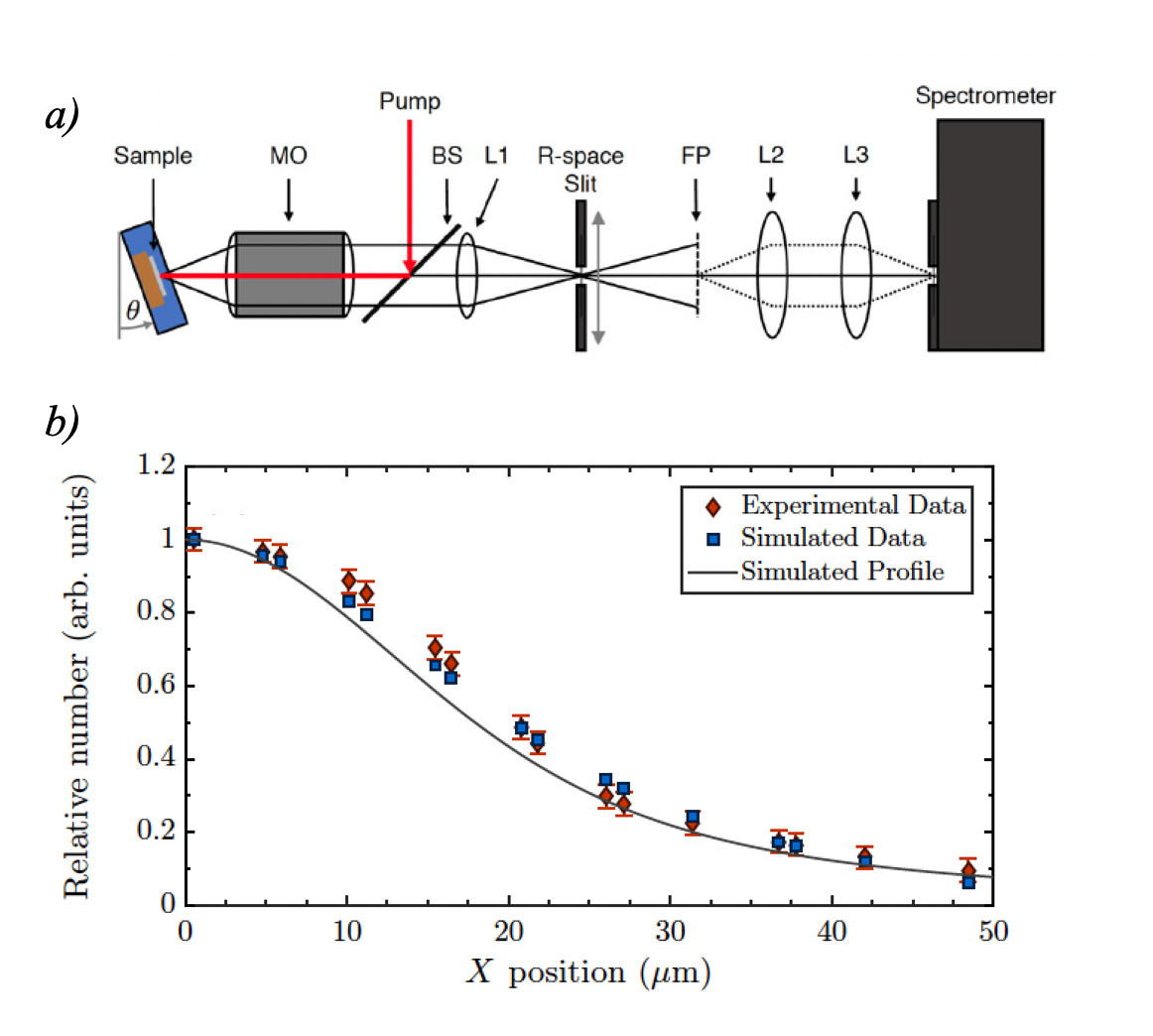}
\caption{a) Apparatus for measuring the emission from exciton-polaritons at high angle, corresponding to large momentum, in the exciton ``bottleneck'' region. b) Diamonds: Measured exciton intensity as a function of distance from a tightly focused laser generation spot. The squares and solid line give the prediction of a simple theory for exciton diffusion. From Ref.~\protect\onlinecite{exdiff}. } 
\label{ex3}
\end{figure}

Ref.~\onlinecite{naturephys} made the assumption that bare excitons stayed within a micron or two of the laser excitation ring, based on earlier measurements of exciton diffusion in quantum wells with no microcavity \cite{QWexdiff}. Similarly, Ref.~\onlinecite{piec} made no direct measurement of the exciton population; for the theory of that paper, the exciton profile at the barriers of the ring was estimated from the blue shift, and then the curve generated from this estimate was extrapolated to the center of the ring.  Subsequent work \cite{exdiff}, however, directly measured the exciton diffusion in the same structure, and found indeed that a significant fraction of the excitons could diffuse 30 microns or more. This experiment was done by tilting the sample (and cryostat) to a steep angle, so that polaritons and excitons with high in-plane momentum $k$ could be observed. This corresponded to directly observing emission from particle momenta up to $5\times10^4$ cm$^{-1}$, which as seen in Figure 1, is well into the excitonic range of the states. As seen in Figure \ref{ex3}, exciton emission was observed at spatial positions well away from the creation point.  This population can arise both from thermal up-scattering from polaritons into exciton states, and from high diffusion constant of the excitons in low-momentum states. Although the larger exciton fraction makes these particles more likely to scatter with the lattice, which would imply lower diffusion constant, this is compensated by the fact that excitons in the ``bottleneck'' region, at the crossover from polariton to exciton character of the states, have the highest group velocity of the whole exciton-polariton band.  


Because of this dark exciton population, and because the interaction of the pure excitons with polaritons is stronger by a factor of 2 at resonance, we can make the assumption that the polariton interactions in this experiment were dominated by their interaction with excitons, and we can neglect entirely the interaction of the polaritons with each other. This at first might seem to make the calibration of the interactions difficult, because the total exciton population density is hard to measure, although it can be estimated both from the high-angle measurements described above and the kinetic numerical simulations discussed in Section IV. However, there is a much tighter constraint provided by these measurements themselves. As seen in Figure~\ref{mit2}(a) and (f), there are {\em two} spectroscopic numbers obtained, namely the blue shift (real part of the self-energy) and the Lorentzian line broadening (imaginary part of the self-energy). As discussed in the Introduction and derived in detail in Ref.~\onlinecite{snokebook2}, the two terms are not independent, but depend on the same polariton-exciton interaction constant $\tilde g$. Since these are first-order and second-order in $\tilde g$, respectively, they have different dependences on $\tilde g$, which then provides two equations for the two unknown values of $\tilde g$ and the exciton density $n_{ex}$. For a discussion of the validity of taking the line broadening data of Ref.~\onlinecite{naturephys} as indicative of the collision rate, see Appendix \ref{appbroad}.

 It is immediately clear from the line broadening data like that shown in Figure~\ref{mit2}(f) that the theoretical number for $\tilde g$ cannot give the line broadening observed, even assuming that the interactions are dominated by the exciton reservoir. An analytical calculation of the scattering rate of polaritons with excitons, derived in Appendix \ref{appscatt}, gives the formula for the Lorentzian line broadening half width,
\begin{eqnarray}
\gamma = \frac{\hbar}{\tau} =  2\pi \tilde{g}^2\frac{m}{\hbar^2}n_{ex},
\label{rate}
\end{eqnarray}
where $m$ is the polariton mass.  If we assume that $\tilde g$ is given by the theoretical value $g_{ex} = 12$~$\mu$eV-$\mu$m$^2$, divided by 12 for the number of quantum wells, and use the measured polariton mass $m = 1.8\times 10^{-4} m_0$ from the curvature of the momentum-space data shown in Figure~\ref{mc1}(c), where $m_0$ is the vacuum electron mass, then this formula implies a Lorentzian line broadening around $10^{-5}$~meV at an exciton density of $10^9$~cm$^{-2}$; the corresponding blue shift would be only $10$~$\mu$eV. Even if the exciton density is assumed to be 50 times larger, so that the blue shift value is in agreement with the experimental value of 0.5 meV, this still only implies a line broadening of  50~$\mu$eV, far below the measured value of 0.75 meV.  Because the line broadening is proportional to $\tilde g^2$ while the blue shift is linear with $\tilde g$, it is not possible to get agreement of the blue shift and line broadening numbers at {\em any} density, even assuming that the interactions are completely dominated by the exciton population, unless $\tilde g$ is increased to around 50-60 $\mu$eV-$\mu$m$^2$. In that case, the experimental values of both the blue shift and the line broadening of around 0.5 meV can be obtained for an exciton density of $10^9$~cm$^{-2}$, consistent with the observed numbers and reasonable estimates of the exciton density, as discussed below. Taking into account the factor of 12 for the MQW structure (which is a debatable approach, as discussed in the introduction) implies a value for $g_{ex} \sim 500~\mu$eV-$\mu$m$^2$ from these measurements, about two orders of magnitude higher than the theoretical value deduced in Section I for this experiment, but well below the originally reported value of Ref.~\onlinecite{naturephys}. As discussed above, if we assume that excitons in nearby quantum wells interact with each other, then this number should be reduced to around 150~$\mu$eV-$\mu$m$^2$, about a factor of ten higher than the theoretical value. 

The above numbers are consistent with the expected ratio of polaritons and excitons in the conditions used in Ref.~\onlinecite{naturephys}. 
The highest polariton density in the data of Ref.~\onlinecite{naturephys} was about $3\times 10^7$~cm$^{-2}$; it could not be much higher, because the onset of Bose-Einstein quantum statistics occurs at around 10$^8$ cm$^{-2}$ for those experimental conditions, giving significant change to the spectral behavior, including line narrowing.  The exciton density of $10^{9}$~cm$^{-2}$, obtained from solving the two equations for $\tilde g$ and $n_{ex}$ as discussed above, therefore implies about 30 excitons per polariton.  This is a reasonable number based on the estimated energy splitting between the polariton ground state and the bare exciton states of $\Delta E \simeq 2$ meV, using the calibration discussed in Appendix~\ref{appcal}. We can estimate that the ratio of excitons to polaritons is given by the Boltzmann factor times the ratio of the densities of states of the two species
\begin{eqnarray}
\frac{n_{x}}{n} = e^{-\Delta E/k_BT}\frac{D_{\rm ex}(E)}{D_{LP}(E)} = e^{-\Delta E/k_BT}\frac{m_{x}}{m} ,
\end{eqnarray}
where $m_{x}$ is the in-plane mass of the bare excitons, equal approximately to $0.2m_0$ \cite{mass}, and $m$ is the lower polariton mass, measured as $1.8\times 10^{-4}m_0$ for this case. Assuming an effective temperature of 10 K gives a Boltzmann factor $ \sim 0.1$, while the ratio of the density of states of the lower polariton and the bare excitons, given by their mass ratio, is of order 1000, which gives a ratio of the two populations around 100 near equilibrium. Numerical simulations of the polariton system like those described in Section~\ref{sect.noneq} and Appendix~\ref{apphart} show that for the case of the long-lifetime structures used in Ref.~\onlinecite{naturephys}, the particles should be near to equilibrium.

Some experiments have used the ``S-curve,'' that is, a plot of the lower polariton population as a function of pump power, to estimate the ratio of excitons to polaritons. At low density in non-resonant excitation experiments, the number of polaritons typically rises linearly with pump power, as the rate of conversion of reservoir excitons into polaritons is nearly constant. At the BEC threshold, the lower polariton population jumps up, because the excitons efficiently convert into polaritons when stimulated down-scattering becomes important. One can therefore take the jump in the polariton population as a lower bound on the number of reservoir excitons before condensation. For short-lifetime systems (e.g., Ref.~\onlinecite{scurve}), the jump can be a factor of several hundred, consistent with the results of numerical simulations described in Section~\ref{sect.noneq} and Appendix~\ref{apphart} for a homogeneous, short-lifetime system.  In the experiments of Ref.~\onlinecite{naturephys}, as in the experiments of Ref.~\onlinecite{myers-pillar}, there is second mechanism for the jump. In these experiments with long polariton lifetime, the pump region was placed far from the region where the polaritons were observed. Above the BEC threshold, the excitons in the pump region convert into polaritons, which can then move quickly into a trap tens of microns distant, so that the jump in density of the polaritons at the BEC threshold will be more than just the amount expected for a homogeneous system. In this case the jump of the S-curve in long-lifetime samples cannot be taken as the ratio of excitons to polaritons, as in the short-lifetime samples.

The revised number of $\tilde{g} \sim 50$ $\mu$eV-$\mu$m$^2$ derived here from the data of Ref.~\onlinecite{naturephys} applies to polaritons at low density, well below the threshold for Bose-Einstein condensation. Two other studies with similar polariton densities have found similar values. One study \cite{walker} found a value of $g \sim 30$ $\mu$eV-$\mu$m$^2$, which when accounting for the exciton fraction and number of quantum wells implies $g_{ex} \sim 200$ $\mu$eV-$\mu$m$^2$, again around two orders of magnitude higher than expected theoretically. In that study, polaritons were allowed to propagate in one dimension in a wave guide, and pulses of different durations were used to account for the accumulation of an exciton reservoir, which was assumed to have lifetime much longer than the polaritons. 

Ferrier et al. \cite{ferrier} found blue shifts and line widths of the polaritons of the order of 0.3 meV when at a density just below the condensation threshold, similar to the results of Ref.~\onlinecite{naturephys} shown in Figure~\ref{mit2}.  They did not attempt to estimate the reservoir exciton density, and reported a value of the exciton interaction only for high densities, well above the condensate threshold, as discussed below.

\section{Spectroscopic measurements of $g$ at high density}
 
 \begin{figure}[b]
\includegraphics[width = .9\linewidth]{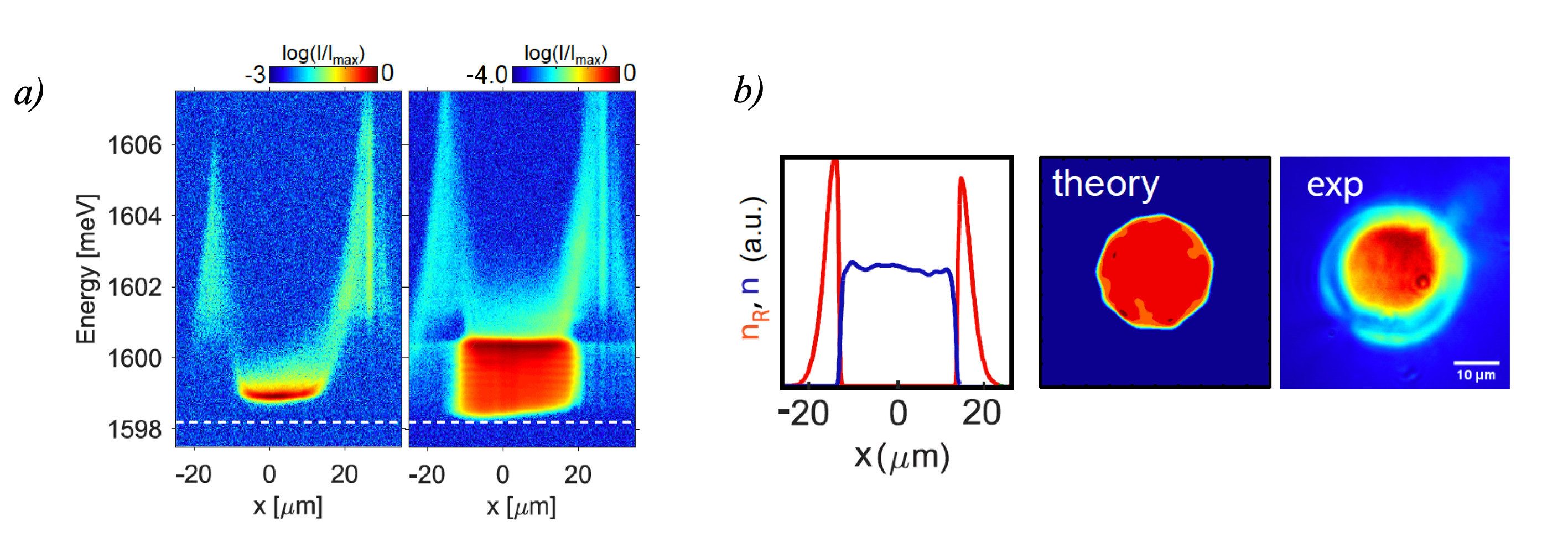}
\caption{a) Spectral images of the emission from a polariton condensate at two different densities, sitting in the midst of an exciton cloud, or reservoir. Note that the polariton condensate in the center has lower energy than polaritons on the sides, where the excitons form a barrier. As the density increases in the image on the right, the ground state energy decreases, while the chemical potential increases. b) Left: plot of the polariton intensity as a function of position in the trap. Right: theoretical and measured intensity profile in the trap.
From Ref.~\protect\onlinecite{ostroTF}. } 
\label{ostro4}
\end{figure}

The large values of of $g_{ex}$ discussed in the previous section were for polaritons at low density. Two other studies using the same structures at high polariton density have found low values of $g$, more consistent with the theoretical value discussed above. One study \cite{ostroTF} used the fact that a polariton condensate stimulates down-conversion from the exciton reservoir. Therefore, at high enough condensate density, the exciton reservoir can be greatly depleted, leading to a strong reduction of the blue shift of the polaritons due to their interaction with the excitons. Figure \ref{ostro4} shows that as the pump intensity is increased, the polariton energy first shifts up as the number of excitons increases, and then red shifts back down toward the original ground state energy as the excitons are increasingly eliminated converted into polaritons by stimulated scattering. The polariton energy then blue shifts again, with a much weaker dependence on density;  at this point the polaritons are in the Thomas-Fermi regime, with a well-defined chemical potential. The value of $g$ extracted in this regime was $0.18~\mu$eV-$\mu$m$^2$, corresponding to $g_{ex} \sim 6~ \mu$eV-$\mu$m$^2$ when the exciton fraction and number of quantum wells are taken into account, comparable to the theoretical value. (As discussed above and in Appendix E, the interaction term is reduced by a factor of 2 in a condensate, but also if the condensate is linearly polarized, then the theoretical value should be multiplied by a factor of two, since spin is no longer random). The same experimental group found a similar value under the same experimental conditions from the slope of the Bogoliubov linear branches \cite{ostrobogo}. Earlier work \cite{brichkin} showed a similar result in the high density condensate regime, when it could be assumed that the dominant interactions were polariton-polariton interactions within the condensate.  The results of Ref.~\onlinecite{ferrier} at high density had much higher values, with $g_{\rm ex}$ over $100~\mu$eV-$\mu$m$^2$ when the number of quantum wells is taken into account, most likely because those experiments were done in tiny structures in which the effect of quantum confinement played a major role. 

\begin{figure}[b]
\includegraphics[width = .96\linewidth]{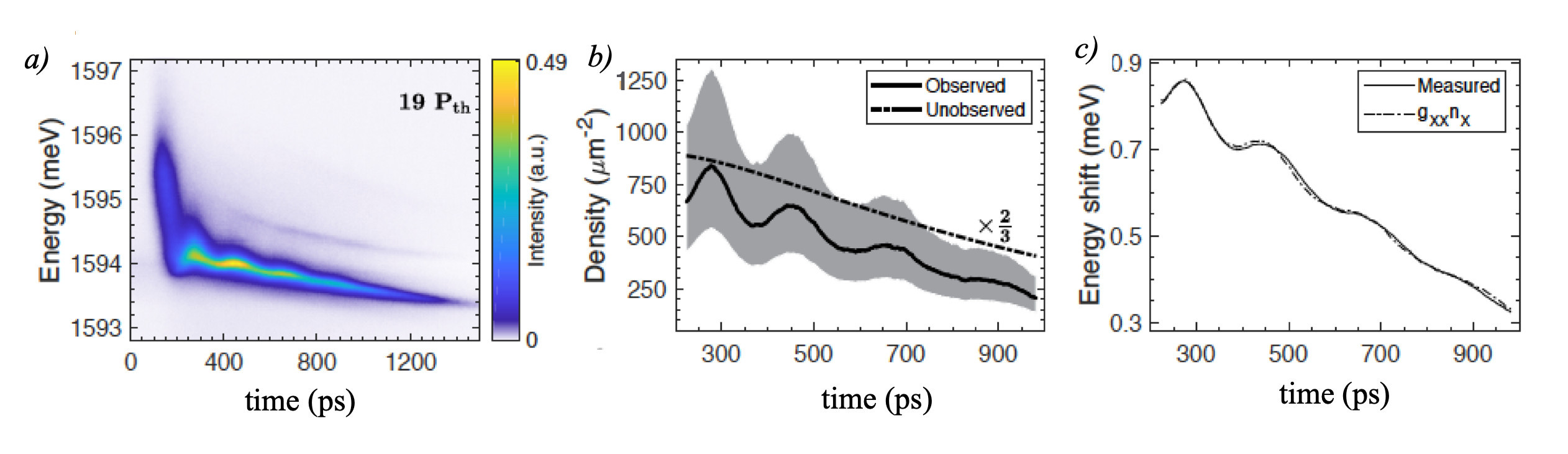}
\caption{a) Time-resolved energy spectrum of a polariton condensate in a one-dimensional trap, recorded with a streak camera. The oscillations arise because the polaritons at late times are effectively in a macroscopic, one-dimensional harmonic potential. b) Solid line: Measured density of the polaritons at the potential energy minimum (gray area gives the uncertainty). Dashed line: estimated exciton reservoir density. c) Comparison of the measured energy shift of the lower polariton line and the predicted energy shift deduced from the density data of (b). The constant of proportionality gives the interaction constant $g$.
From Ref.~\protect\onlinecite{ring}. } 
\label{shou5}
\end{figure}
Another recent study \cite{ring} used one-dimensional wires etched from the same original III-V structures as used in Refs.~\onlinecite{naturephys} and \onlinecite{ostroTF}, and observed oscillations of the polariton population due to the interplay between a potential energy gradient (caused by a wedge in the cavity thickness) and the kinetic energy of the polaritons. The oscillations in intensity of the polaritons were measured, which could then be calibrated to an absolute density oscillation, and this was compared to measurements of the oscillation of the polariton energy position, as shown in Figure \ref{shou5}.  This experiment gave a stringent measurement of the polariton-polariton interaction strength, because it could be assumed that only the polaritons oscillate---the excitons in the system have mass three orders of magnitude larger, which for this geometry would give negligible oscillations of the exciton reservoir. Therefore the polariton contribution to the signal can be separated from the slowly-varying exciton contribution. The analysis of this experiment, which was done at high density in the condensate regime, also gave a value for the interaction strength very close to the theoretical value, namely $g_{ex} = 12 \pm 6$ $\mu$eV-$\mu$m$^2$. 
There are therefore two different types of experiments done on the same structures that give a value of the polariton-polariton interaction parameter up to two orders of magnitude lower than the the value at low density, discussed in Section II, even when all the effects of an exciton reservoir are taken into account. 

As discussed in Appendix~\ref{appscreen}, a hard-sphere boson system has a reduction of the effective interaction of a factor of 2 when in the quantum-degenerate regime, but many-body physics of bosons alone is not expected to give a greater reduction than that.  It may be that disorder plays an important role of enhancing the effective interaction at low density; for example, if the spatial correlation of the excitons is not uniform, but instead involves them being clustered near each other in local minima. The effects of disorder are reduced when there are extended coherent states, or when there are only single states involved, as in the case of micro-photoluminescence from quantum dots. 

\section{Estimates of the interaction strength from nonequilibrium measurements}
\label{sect.noneq}

The above issues lead us to revisit earlier results from short-lifetime structures of the same type of III-V structure. A large number of early experiments were done with structures that had a cavity photon lifetime of around 1 ps \cite{yamamoto,science,bloch}. The only significant difference between those early samples and the modern samples with cavity photon lifetime of 100 ps (and therefore polariton lifetime at resonance of around 200 ps) is the number of periods of the distributed Bragg reflectors (DBRs) used to make the mirrors of the cavities.  The GaAs quantum wells in both types of sample were the same.

A numerical study \cite{hartwell} fit the nonequilibrium BEC distribution using a quantum Boltzmann equation to model the scattering dynamics.  This model was sophisticated, treating the exciton and polariton populations as one continuous band, rather than as two distinct populations, and taking into account all of the phonon emission processes, screened-Coulomb scattering with free electrons, and the momentum dependence of the exciton fraction in the polariton-polariton scattering cross section. The steady-state distribution of the particles was calculated by evolving their distribution in time until the distribution no longer changed. This did not give equilibrium distributions with a well-defined temperature, but rather nonequilibrium distributions in steady state, in which the particle losses due to photon loss equalled the number generated by the pump laser, which fit the experimentally measured distributions well.

The nonequilibrium nature of the data from these early experiments actually gives a strong constraint on the polariton-polariton interaction strength that equilibrium distributions do not give.  In equilibrium (seen e.g. in Ref. \onlinecite{equilprl}), the distribution depends only on the chemical potential and the temperature, and is independent of the details of the scattering processes. When the lifetime is short, however, comparable to the scattering time, then the shape of the nonequilibrium distribution is highly sensitive to the exact details of the scattering cross section. 

In the experiments reported in Ref.~\onlinecite{hartwell}, no measurement was made of the absolute polariton density. Instead, the theoretical value of Ref.~\onlinecite{tassone} for the polariton-polariton scattering cross section was assumed, and the density was varied in the calculations to get a fit. In addition, the assumption that the theoretical interaction strength must be divided by the number of quantum wells was not used, and so the value used was nominally 12 times too large; i.e., the cross section was 144 times larger than the actual theory prediction. 

Good fits were obtained, but even with the cross section two orders of magnitude larger than the theoretical value, the densities needed to get fits were well above what we now know to be typical densities for the polaritons at the BEC threshold. We have updated the numerical model of Ref.~\onlinecite{hartwell} to include a second population of dark excitons which couple to the bright exciton-polariton band by phonon emission and absorption and to account for realistic conditions of lattice temperature and background free charge (see Appendix \ref{apphart}). Figure \ref{hart6} shows the results of these new calculations. In each case,
we have fixed the interaction strength $g$ and increased the pumping density until occupation numbers greater than 1 appear in steady state, that is, to the point when Bose-Einstein statistics become important. A crucial element of this model is that it completely describes the entirety of the exciton and polariton populations; the exciton reservoir is fully taken into account by the higher-energy exciton-polariton and dark exciton states.

\begin{figure}
\includegraphics[width = .65\linewidth]{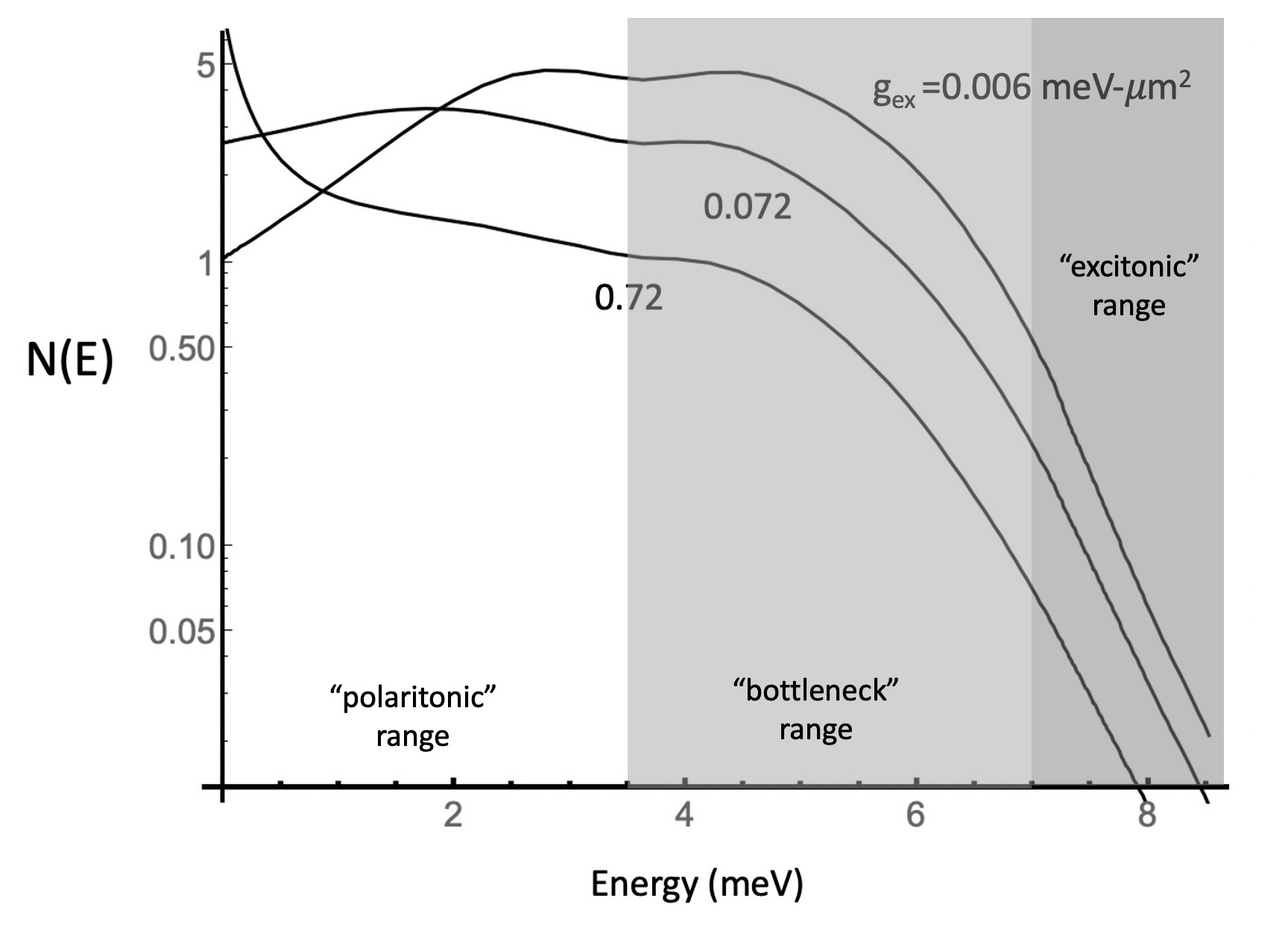}
\caption{a) Results of the numerical simulation using a quantum Boltzmann equation for the steady-state energy distribution of a gas of polaritons for a spatially homogeneous polariton population, for three different interaction strengths.  Continuous pumping of the exciton population (i.e., states above 7 meV energy) was used to maintain a steady-state polariton distribution. The labels of the curves give the exciton-exciton interaction $g_{ex}$ in units of meV-$\mu$m$^2$. 
For the three curves, the case of  $g_{ex}=0.006$ corresponds to a polariton density of $1.8\times 10^7$ cm$^{-2}$ and total exciton density (dark and bright) of $2.6\times 10^{10}$ cm$^{-2}$; the case of $g_{ex} = 0.072$ corresponds to a polariton density $4.2\times 10^7$ cm$^{-2}$ and total exciton density $1.7 \times 10^9$ cm$^{-2}$; the case of $g_{ex} = 0.72$ corresponds to a polariton density $5.9 \times 10^7$ cm$^{-2}$, total exciton density $8.0\times 10^9$ cm$^{-2}$  All three plots used the following parameters: cavity photon lifetime of 1.2 ps; division of $g_{\rm ex}$ by 12 to account for 12 quantum wells; exciton and cavity modes resonant at $k = 0$; lattice temperature of 15 K; phonon emission and screened scattering with a small free electron population ($4\times 10^8$ cm$^{-2}$) were accounted for as in Ref.~\protect\cite{hartwell}, and dark excitons coupled to the exciton-polariton band were incorporated as discussed in the text and Appendix~\ref{apphart}.} 
\label{hart6}
\end{figure}
Figure \ref{hart6} shows the steady-state energy distribution generated for three different exciton-exciton interaction strengths $g_{ex}$, for a cavity photon lifetime of 1.2 ps and other parameters similar to those of the experiments of Ref.~\onlinecite{science} and other similar experiments with short lifetime cavities. When $g_{ex}$ is equal to 6~$\mu$eV-$\mu$m$^2$, the distribution function is completely different from the results of Ref.~\onlinecite{science} and other related work (e.g. Refs.~\onlinecite{yamamoto} and \onlinecite{bloch}), which showed a peak in the ground state and a quasi-thermal tail to higher energy. If we keep $g_{ex}$ at this value and increase the pumping, there is no fit to the experimental data at {\em any} density; instead, the exciton states at much higher energy than the polariton ground state become highly occupied. This can be understood by a simple estimate: the line broadening of Equation (\ref{rate}) equal to $3\times 10^{-5}$ meV at a density of $10^9$ cm$^{-2}$ calculated in Section II for the theoretical value of $g_{ex}$ corresponds to a scattering time of $\tau = \hbar/\gamma = 22$ ns, far longer than the polariton lifetime of  200 ps, and presumably even longer than the intrinsic exciton lifetime. The interactions of the excitons are simply too weak for them to scatter down into polariton states, even when stimulated scattering into the polariton states is taken into account. If the density is increased to make this particle-particle scattering time shorter, the occupation number in excitonic range of the dispersion will exceed the quantum degenerate threshold long before their scattering is fast enough to give significant scattering into polariton states, leading to nonequilibrium condensation in the excitonic states instead of in the polariton ground state. 
 
 Figures \ref{hart6} shows the same model when $g_{ex}$ is increased by factors of 10 and 100.  For the case of $g_{ex}$ increased by a factor of 10 shown in (b), once again the occupation numbers greater than 1 appear first in excitonic states at high energy.  Only for the highest value of $g_{ex}$ is the curve qualitatively the same as the experimental data. This is precisely the range of $g_{ex}$ which agrees with the line broadening data discussed in Section II. There is therefore consistency in the numerical modeling fits of Ref.~\onlinecite{hartwell} as updated here, and the blue shift and the line broadening data of Ref.~\onlinecite{naturephys}.
 
Because the scattering rate is proportional to the square of the interaction energy, the nonequilibrium data discussed here provide a much more sensitive measure of the value of $g_{ex}$. Any analysis that presumes that the exciton-exciton interaction strength is as low as the nominal theoretical value must have an explanation for how so many experiments with short polariton lifetime can show any thermalization or quasi-thermalization at all, when the theoretical interaction strength gives such long scattering times. 

\section{Other estimates of the polariton-polariton interaction strength}

 In Ref.~\onlinecite{Delteil2019}, the temporal correlation of photons emitted from polaritons was measured, which according to theoretical calculations gives an indirect measurement of the polariton-polariton interaction, because the polaritons repel each other, making it less likely for two to be in the same place at the same time. A single quantum well was embedded in a cavity in order to maximize the polariton-polariton repulsion; the Rabi splitting between the upper and lower polariton states was 3.5 meV, compared to values of 14-15 meV in the structures discussed above.

Theory \cite{block1,block2} for polariton-polariton correlations predicts that the maximum dip in $g^{(2)}(\tau = 0)$ (where $g^{(2)} = 1$ is expected for completely uncorrelated particles) should be approximately $A = g/S\gamma$, where $g$ and $\gamma$ are the polariton-polariton interaction strength and line width, as defined above, and $S$ is the observed area in the measurement. For the observed value of $A \sim 4\%$, the observed area $S = 3$ $\mu$m$^2$, the measured line width of 0.05 meV, and a correction factor of about a factor of 2 for nonequilibrium effects, this implied $g_{ex} \sim 40$ $\mu$eV-$\mu$m$^2$, about a factor of four higher than the theoretical value. The densities used were about an order of magnitude below the threshold for onset of coherence in this system. In general, in systems with one or just a few quantum wells, Bose condensation of the polaritons is not expected, and the polaritons enter coherence by standard lasing \cite{boser}, and the density for this threshold may be higher than the threshold for BEC \cite{jap}.   A similar experiment was reported in Ref.~\onlinecite{Matutano2019}, using a single InGaAs quantum well in a microcavity, and found an experimental value of $g_{ex}$ in the range of 3-6 times the expected theoretical value. No polariton density was reported for this experiment, but it is likely that it was done at a density similar to that of Ref.~\onlinecite{Delteil2019}. 


Another work \cite{bloch2017} used the nonlinearity of the optical response of a polariton system to deduce the ratio of $g$ to the cavity line width, and reported a value for the interaction strength corresponding to $g_{ex} = 30$ $\mu$eV-$\mu$m$^2$,which is above the theoretical value, but not by much. Since these experiments involved resonant excitation, these experiments may resemble more closely the results at high density with a coherent condensate.

Table~\ref{gtable} gives a summary of the values for $\tilde g$ for low density, incoherent polariton populations.

\begin{table}
\begin{tabular}{|c|c|c|c|c|c|}
\hline
\hspace{.2cm} $\tilde g$ ($\mu$eV-$\mu$m$^2$) \hspace{.2cm}  & \hspace{.2cm}  exciton fraction \hspace{.2cm}   & \hspace{.3cm} $\hbar\Omega$ (meV)   \hspace{.3cm}  & \hspace{.1cm}  $N_{QW}$ \hspace{.1cm} &  \hspace{.3cm} QW material  \hspace{.3cm} &  \hspace{.3cm} references  \hspace{.3cm}  \\ \hline
50 & 85\% & 14 &  12 & GaAs & \protect\cite{naturephys}, this work \\ 
20  & 60\% &  3.5& 1 & InGaAs &  \protect\cite{Delteil2019} \\
40 & 60\% & 9 & 3 & GaAs  & \protect\cite{walker} \\
$50^*$  & 50\% & 15  & 12 & GaAs  & \protect\cite{ferrier}\\
\hline
\end{tabular}
\caption{Estimated values of the polariton-exciton interaction strength $\tilde g$ in the low density (non-condensate) limit; $N_{QW}$ is the number of quantum wells, and $\hbar\Omega$ is the Rabi splitting of the upper and lower polariton branches. In cases where a polariton-polariton interaction $g =|\alpha|^4 g_{\rm ex}$ was reported, this has been adjusted to $\tilde g = g/ |\alpha|^2$.\\
$^*$Value computed from the reported blue shift and line width data in the low density limit, using the method discussed in Section II.
}
\label{gtable}
\end{table}

\section{Conclusions}

After various corrections have been taken into account, the value of the exciton-exciton interaction constant in GaAs/Al$_x$Ga$_{1-x}$As microcavity structures from experiments at low polariton density ranges from $g_{ex} \sim 30$~$\mu$eV-$\mu$m$^2$ to $g_{ex} \sim 500$~$\mu$eV-$\mu$m$^2$, with all values lying above the theoretical value.  As discussed above, lower values of the interaction strength fail by orders of magnitude to give a thermalization time of the polaritons adequate to form a condensate within their lifetime, because the thermalization rate is proportional to the square of the interaction strength, and therefore drops off rapidly for low values of $g$. 

In absolute terms, the measured values of the exciton-polariton interaction strength $\tilde g$ in structures with multiple quantum wells all fall in the same range as those of single quantum wells, namely, $\tilde g \sim 50 ~\mu$eV-$\mu$m$^2$. The high values of $g_{ex}$ taken from measurements in MQW structures come from multiplying the measured values of $\tilde g$ by the number of quantum wells. If the excitons interact efficiently with each other in the vertical direction, then a lower correction term should be used, which would give much less discrepancy. As discussed in Section I, multiplying the measured $\tilde g$ by the full number of quantum wells to get $g_{ex}$ is based on the assumption that excitons in different quantum wells are non-interacting. In typical structures, however, some quantum wells can be separated only by distances of the order of the excitonic Bohr radius, and therefore one expects excitons to interact with other excitons in nearby wells. For the specific case of Refs.~\onlinecite{naturephys} and \onlinecite{ferrier}, a quantum well structure was used with three groups of 4 quantum wells. If we assume excitons within a group of four wells do interact, while those in different groups do not interact, this gives a correction factor of 3 rather than 12. In this case, the experiments are consistent with $g_{\rm ex}$ $\sim$ 150 $\mu$eV-$\mu$m$^2$, compared to the theoretical expectation of 12-15 $\mu$eV-$\mu$m$^2$. 

It is also not absolutely clear that the simple assumption made in the Introduction, that the polariton-polariton interaction strength is just given by the exciton-exciton interaction strength times the exciton fraction of the polaritons. The two-level coupling model for the polaritons is known to break down when the width of the LP and UP lines is comparable to their separation; a Feshbach-type resonance with the biexciton states can also complicate the calculation \cite{fesh}.

Measurements when there are coherent extended states, as in BEC, consistently give values of the interaction constant that are lower than the values deduced at low density, and in the range predicted by the theory. In general, the many-body effects of screening and anticorrelation are expected to reduce the effective interaction strength and the blue shift from the nominal value \cite{zimm,laikht}, but the values reported at low density here are still greater than expected, and may be an effect of disorder in these structures. For example, as suggested by Keeling \cite{keeling-private}, polaritons may tend to congregate together in energy minima, causing them to feel an effectively higher density. This would not by itself explain why the scattering rates increase even more strongly, since they are propotional to $\tilde g^2 n$ while the blue shift is proportion to $\tilde g n$, but it may be part of the explanation. This effect would not occur for ensembles of quantum dots, because each particle is isolated in its own dot, so that spatial correlation effects will not occur.

{\bf Acknowledgements}. This work has been supported by the National Science Foundation through grant DMR-2004570. 
We thank Jonathan Keeling and Sandy Fetter for useful discussions.

\appendix

\section{Line broadening in polariton systems}
\label{appbroad}

The deduction of $g$ from the comparison of blue shift and line broadening data raises the general question of how to understand line broadening data in general in polariton systems.   The Lorentzian line broadening invoked above, giving rise to Equation (\ref{rate}), is known as ``homogeneous'' broadening, as well as ``lifetime'' broadening or ``collision broadening.
This type of line broadening can be understood physically as an implication of the uncertainty principle $\Delta E \ge \hbar/\Delta t$. When the time spent by a particle in a given state is $\Delta t$, the energy of that state cannot be defined more narrowly than $\hbar/\Delta t$. As discussed in Section I, this type of line broadening arises from an imaginary self-energy in quantum mechanical theory; Ref.~\onlinecite{snokebook2}, Section 8.4, shows that an imaginary self-energy gives rise to a Lorentzian line shape. 

The total homogeneous linewidth will be proportional to the sum of all out-scattering rates, i.e.,
\begin{eqnarray}
\gamma = \hbar\left( \frac{1}{\tau_R} + \frac{1}{\tau_{\rm scatt}} + \frac{1}{\tau_{\rm phon}} \right),
\end{eqnarray}
where $\tau_R$ is the radiative lifetime, $\tau_{\rm scatt}$ is the particle-particle scattering rate deduced above, and $\tau_{\rm phon}$ is the particle-phonon scattering rate, if any. It has sometimes been claimed that the line width is a direct measure of the radiative lifetime, but this is only the case when the lifetime is so short that $1/\tau_R$ is larger than all other terms. When the lifetime is long, the broadening due to lifetime can become negligible compared to other terms. For example, a lifetime of 200 ps or so, as in the experiments of Ref.~\onlinecite{naturephys}, gives only $\hbar/\tau_R \simeq $ 3 $\mu$eV, well below the resolution of most spectrometers.  Exciton-phonon interaction times at low temperature are also very low, nanoseconds or longer, so that the phonon term typically gives an even smaller contribution to the lower polariton broadening, although phonon broadening of the {\em upper} polariton state can be significant since there is a direct channel to convert into lower polaritons by phonon emission. This also explains why the photoluminescence from the upper polaritons is much weaker in high-Q microcavities: in these structures, the rate of photon emission is much lower than the rate of phonon emission to drop down into the lower polariton states.

In addition to homogeneous Lorentzian line broadening, there are also several types of ``inhomogeneous'' line broadening, which must be convolved with the homogeneous line broadening.  Inhomogeneous broadening arises from integration of signal over a range of energies. This can come about due to several sources. In semiconductors, there is typically alloy and impurity disorder and quantum well width disorder that give fluctuation of the band gap. These fluctuations typically have a Gaussian distribution, as expected from the central limit theorem for random fluctuations. As seen in Figure \ref{mit2}(f), the lineshape of the polaritons is well fit by a Voigt profile, which is a convolution of a Gaussian and a Lorentzian. The Gaussian term is presumably due to inhomogeneous disorder and is independent of the polariton density, while the Lorentzian part has a width proportional to the polariton density.

In addition, there can be a systematic variation of the potential energy in the area of observation of the photoluminescence. For example, in the polariton structures discussed here, there is almost always a wedge in the cavity thickness. For a typical gradient of 5 meV/mm in the Princeton samples used in Ref.~\onlinecite{naturephys}, an excitation spot size of 10 $\mu$m will have a variation of 50 $\mu$eV, comparable to broadening due to the radiative lifetime given above. There can also be a gradient of the exciton cloud density. In Ref.~\onlinecite{naturephys}, this acted to cancel the effect of the wedge due to the cavity because as the particle density increased, the particles moved to find a common chemical potential, leading to flattening of the potential to less than 10 $\mu$eV over a 10 $\mu$m range.

There can also be temporal variation of the potential energy due to fluctuations of the pump laser intensity, which then give a variation in the exciton reservoir density. If the signal is integrated over a time scale long compared to these fluctuations, this can give substantial inhomogeneous broadening. Much early work had line broadening of this type, until work by Love and coworkers \cite{skolT2} showed the importance of using a stabilized laser. All of the recent work on steady-state populations of polaritons (e.g. Refs. \onlinecite{naturephys} and \onlinecite{exdiff}) used a stabilized M-Squared laser to make this effect negligible.

There are two other, more subtle types of inhomogeneous broadening, as well. One comes from the fact that in all of the photoluminescence meaasurements, there is a range of angles of light emission collected by a lens.  Because there is dispersion of the lower polariton energy, as illustrated in Figure \ref{mc1}(a), this will give a range of energies collected by the system. A 5$^\circ$ external collection angle corresponds to around 0.5 meV of energy range in the parabolic dispersion curve. For the line broadening data reported in Ref.~\onlinecite{naturephys}, angle-resolved data were used to obtain the line width at $k_{\|} = 0$, with angular resolution of approximately 0.5$^\circ$, corresponding a 5 $\mu$eV polariton energy range. 

In a trapped system, there is also the possibility that multiple quantized trapped states are occupied, as seen, e.g., in Ref.~\onlinecite{piec}. If the spacing between these is large, as in a small trap, then these can each be separately resolved, each with its own line width (see, e.g., Figure 1(b) of Ref.~\onlinecite{shouvik}). If the spectrometer resolution is not high, however, then these states can be integrated together to give a broadened line. This will only be significant if the higher trapped states have significant $k_{\|} = 0$ component, which occurs in a small trap; in a large trap, the states will resemble plane-wave states which are separately resolved by angle-resolved imaging. The diffraction limit implies that the finite spatial dimension $L$ for any measurement gives a resolution limit $\Delta k_{\|} \sim 1/L$, which means that confined states are smeared in $k$-space by this amount.

Our conclusion for the line width data of Ref.~\onlinecite{naturephys}, taking all of these effects into account, is that at high density, but still below the condensation threshold density, the Lorentzian line broadening arises from the collisional process controlled by polariton-exciton scattering. None of the homogeneous broadening processes discussed here are expected to give a Lorentzian line shape with density-dependent width as measured here.

In the condensate regime, when the occupation number of the ground state is larger than 1, spectral narrowing will occur. (See Ref.~\onlinecite{snokebook2}, Section 9.5 for a derivation.) This has been seen in all of the polariton condensate experiments and is typically used as a telltale for condensation.

\section{Derivation of the rate of polariton-exciton scattering}
\label{appscatt}

The out-scattering rate per particle in a momentum state $\vec{k}$ in a two-dimensional system is written as (cf. Ref.~\onlinecite{snokebook2}, Chapter 4)
\begin{eqnarray}
\frac{1}{N_k}\frac{\partial N_k}{\partial t} &=& \frac{2\pi}{\hbar}  \sum_{p,q} 4U^2 N_p \delta(E_k+E_p - E_q- E_{\vec{k}+\vec{p}-\vec{q}}) \nonumber\\
&=& \frac{2\pi}{\hbar} \left(\frac{4(UA)^2}{(2\pi)^4} \right) \int d^2q \int_0^{\infty} p dp \int_0^{2\pi} d\theta   N_p \delta(E_k+E_p - E_q- E_{\vec{k}+\vec{p}-\vec{q}}) \nonumber\\
&=& \frac{2\pi}{\hbar}  \frac{g^2}{(2\pi)^4} \int d^2q \int_0^{\infty} N_p  p dp \ 2\int_{-1}^{1} d(\cos\theta) \frac{1}{\sin\theta}   \delta(E_k+E_p - E_q- E_{\vec{k}+\vec{p}-\vec{q}}), \nonumber\\
\end{eqnarray}
where the sum over $\vec{p}$ is for all possible incoming excitons, and the sum over $\vec{q}$ is for the outgoing exciton in each collision.
Here we use $g = 2U/A$, where $U$ is bare interaction and $A$ is the area. As discussed in Section I, boson exchange at low density implies a factor of 2 larger effective interaction.
To resolve the $\delta$-function, when $\vec{k}$ and $\vec{k}+\vec{p}-\vec{q}$ correspond to the incoming and outgoing polariton with mass $m$, respectively, and $\vec{p}$ and $\vec{q}$ correspond to the incoming and outgoing exciton with mass $m_x$, we compute
\begin{eqnarray}
E' = E_k+E_p - E_q- E_{\vec{k}+\vec{p}-\vec{q}} = \frac{\hbar^2}{2m}(k^2 -  |\vec{q}-\vec{k}|^2 - 2|\vec{q}-\vec{k}|p\cos\theta - p^2) + \frac{\hbar^2}{2m_x} (p^2  - q^2),  \nonumber
\end{eqnarray}
which then implies
\begin{eqnarray}
\frac{\partial E'}{\partial (\cos\theta)} = - \frac{\hbar^2}{m} |\vec{q}-\vec{k}|p .
\end{eqnarray}
Then changing variables from $\cos\theta$ to $E'$ and using the $\delta$-function to eliminate $E'$ gives
\begin{eqnarray}
\frac{1}{N_k}\frac{\partial N_k}{\partial t} &=& \frac{2\pi}{\hbar}  \frac{g^2}{(2\pi)^4}\int d^2q  \int_0^{\infty} N_p p dp \  \ 2\int_{-1}^{1} dx \frac{1}{\sqrt{1-x^2}}   \delta(E_k+E_p - E_q- E_{\vec{k}+\vec{p}-\vec{q}}) \nonumber\\
&=& \frac{2\pi}{\hbar}  \frac{g^2}{(2\pi)^4} \frac{m}{\hbar^2}\int d^2q \int_0^{\infty}  N_p  dp   \frac{1}{|\vec{q}-\vec{k}|} \frac{1}{\sqrt{1-x_0^2}}   \Theta(x_0+1)\Theta(1-x_0), \nonumber
\end{eqnarray}
with
\begin{eqnarray}
x_0 = \frac{(k^2-q'^2)+(m/m_x)(p^2-q^2)}{2q' p} .
\end{eqnarray}
In the limit $m_x \gg m$, this becomes
\begin{eqnarray}
x_0 = \frac{(k^2-q'^2)}{2q' p} \equiv \frac{A}{p} .
\end{eqnarray}

We apply the limits of $x_0$ to the integration over $p$. This gives
\begin{eqnarray}
\frac{1}{N_k}\frac{\partial N_k}{\partial t} &=&  \frac{2\pi}{\hbar}  \frac{g^2}{(2\pi)^4} \frac{2m}{\hbar^2}  \int d^2q  \int_{|A|}^{\infty}  N_p  dp  \frac{1}{|\vec{q}-\vec{k}|} \frac{1}{\sqrt{1-(A/p)^2}} .   
\end{eqnarray}
We then change the variable of integration to $\vec{q}' = \vec{q}-\vec{k}$. Then 
\begin{eqnarray}
A =  \frac{k^2- q'^2}{2q' }.
\end{eqnarray}
Then since there is no $\theta'$-dependence, we have
\begin{eqnarray}
\frac{1}{N_k}\frac{\partial N_k}{\partial t} &=&    \frac{g^2}{(2\pi)^2} \frac{2m}{\hbar^3}  \int dq'  \int_{|A|}^{\infty}  N_p  dp  \frac{1}{\sqrt{1-(A/p)^2}} .   
\end{eqnarray}
with $N_p = e^{-\hbar^2p^2/2m_xk_BT}e^{\mu/k_BT}$.
The integrals can be done analytically, to give
\begin{eqnarray}
\frac{1}{N_k}\frac{\partial N_k}{\partial t} &=&    \frac{g^2}{(2\pi)^2} \frac{2m}{\hbar^3}  \int dq'  N_{(k^2-q'^2)/q'}   \sqrt{\frac{\pi m k_BT}{2\hbar^2}} \nonumber\\
&=&  \frac{g^2}{(2\pi)^2} \frac{2m}{\hbar^3}  \frac{\pi m_x k_BT}{\hbar^2}e^{\mu/k_BT}.
\end{eqnarray}
We also have 
\begin{eqnarray}
N_x &=& \frac{A}{(2\pi)^2} \int_0^\infty 2\pi p dp N_p =    \frac{A}{(2\pi)^2} \int_0^\infty 2\pi p dp \  e^{-\hbar^2p^2/2mk_BT}e^{\mu/k_BT} \nonumber\\
&=&  \frac{A}{(2\pi)^2} e^{\mu/k_BT} \frac{m_xk_bT}{\hbar^2}
\end{eqnarray}
so that
\begin{eqnarray}
\frac{1}{N_k}\frac{\partial N_k}{\partial t} &=&      \frac{g^2 m}{\hbar^3} 2\pi  \left(\frac{A}{(2\pi)^2} \frac{ m_x k_B T}{\hbar^2} e^{\mu/k_BT}\right).\nonumber\\
&=& \frac{2\pi}{\hbar}  \frac{g^2 m}{\hbar^2}  \frac{N_x}{A}.
\end{eqnarray}
This is of the order of magnitude expected from unit analysis, namely from Fermi's golden rule, using $2\pi/\hbar$ times the matrix element squared and the density of states for a two-dimensional system, ${\cal D}(E) dE = m/2\pi\hbar^2$. 

The same result is obtained if $\vec{k}$ and $\vec{q}$ correspond to the incoming and outgoing polariton with mass $m$, respectively, and $\vec{p}$ and $\vec{k}+\vec{p}-\vec{q}$ correspond to the incoming and outgoing exciton with mass $m_x$. In this case the heavy-exciton-mass approximation means that $k \ll p$ is assumed.

Note that in this calculation there are no divergences that need to be addressed, and no use of any cutoffs in the integrals. These sometimes arise because the scattering rate is cast in terms of a scattering cross section $\sigma$, using $1/\tau = n\sigma \bar{v}$, where $n$ is the density and $\bar{v}$ is an average velocity. Since the scattering rate derived here is not proportional to $\bar{v}$, one might take this as implying that the cross section $\sigma$ is proportional to $1/\bar{v}$, that is, proportional to $1/k$, which has an infrared divergence. However, it is more appropriate to take the scattering rate as proportional to the density of states in all cases, and to note that $D(E) \propto k \propto \bar{v}$ in three dimensions, while in two dimensions, the density of states is a constant. 

\section{Calibration of the exciton and photon fractions in long-lifetime microcavity structures}
\label{appcal}

In short-lifetime microcavity structures, the exact detuning and exciton fraction of the polariton dispersion can be found easily at every point in the sample by measuring both the photoluminescence (PL) and reflectivity spectra of the upper and lower polaritons. In the long-lifetime structures developed for the experiments discussed in Sections II and III (discussed at length in Refs.~\onlinecite{prx} and \onlinecite{steger-prb}), neither of these measurements is easily possible. Reflectivity is made difficult because the mirrors in these structures have reflectivity greater than 0.9999. As a consequence, the linewidths of the dips in the reflectivity are too narrow to resolve. At the same time, PL from the upper polariton is nearly impossible to see. This is because polaritons in the upper state can scatter into the lower polariton states by efficient phonon emission. While this rate is essentially the same for upper polaritons in both short-lifetime and long-lifetime structures, the rate of radiative emission is around a factor of 100 lower in the long-lifetime structures. The branching ratio of these two avenues of depletion of the upper polariton state therefore implies PL from the upper polariton branch that is 100 times weaker. 

\begin{figure}[b]
\includegraphics[width=0.9\textwidth]{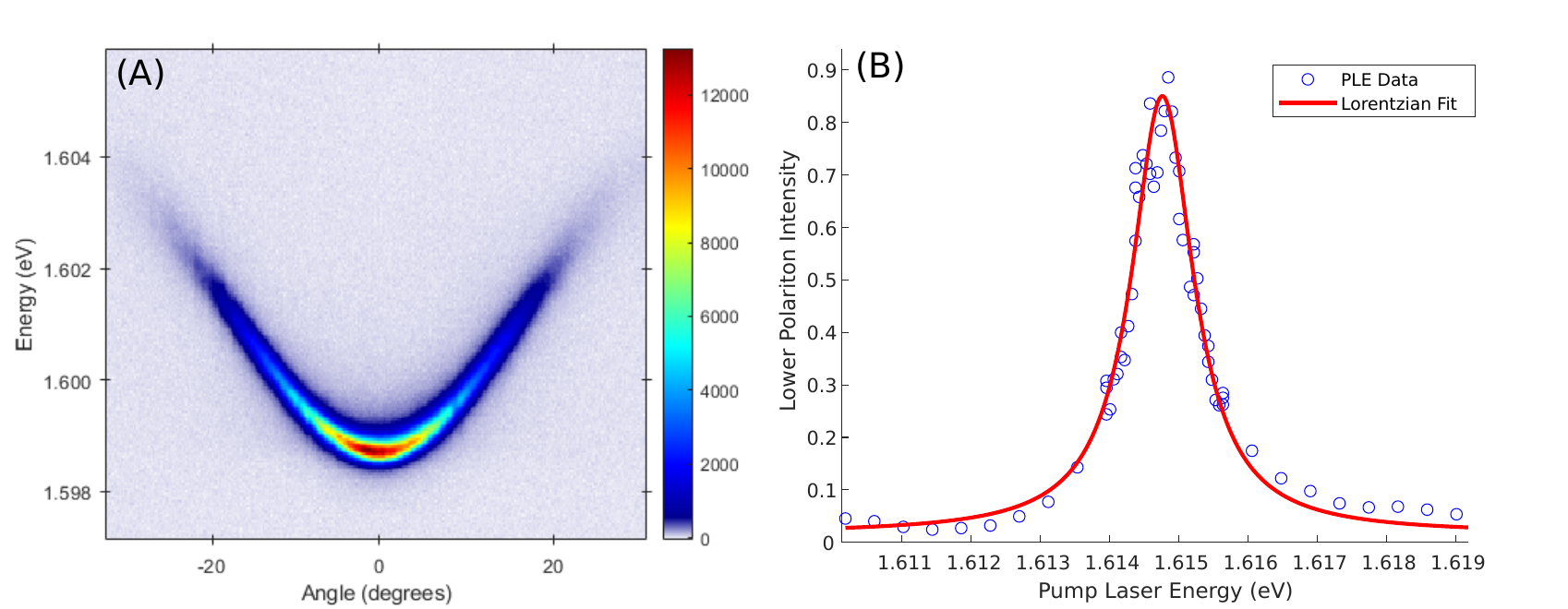}
\caption{(a) Typical angle-resolved lower polariton luminescence from a long-lifetime polariton microcavity structure, created by pumping directly into the upper polariton.  Because the pump laser is only $\sim$20 meV away, care must be taken to properly handle background subtraction.  The PL in this image is integrated over angle and divided by the pump laser's power to give the relative intensity of the lower polariton.  (b) A typical PLE sweep to measure the upper polariton.  The data is fit with a Lorentzian to determine a value for the upper polariton at $k_{\|} = 0$. This measurement is performed at many different locations (detunings) on the sample. Data for Princeton sample 4-6-15.1, used in Refs.~\onlinecite{myers-pillar}, \onlinecite{ostroTF}, and \onlinecite{naturephys}.}
\label{cal1}
\end{figure} 

To overcome this, we used photoluminescence excitation spectroscopy (PLE) to identify the upper polariton state. Typical results are shown in Figure~\ref{cal1}. 
Although the upper polariton state is clearly identified, it is also quite broad, due to the line broadening discussed in Section II, which is much stronger for the upper polaritons than for the lower polaritons due to their much greater phonon emission rate. 

With the lower polariton dispersion of Figure \ref{cal1}(a) and the upper polariton measurement at normal incidence of Figure~\ref{cal1}(b), it is tempting to attempt to fit this data using a simple two-state model. However, this model quickly stops being simple. When complex energies are utilized to include the effects of line width into such a model, assuming the cavity line width is negligible (a good assumption in high-$Q$ cavities), at a resonant detuning such that the exciton energy $E_x$ equals the cavity photon energy $E_c$, the lower polariton's energy would be
\begin{eqnarray}
E_{lp} = E_x - \frac{i}{2}\gamma_x - \frac{1}{2}\sqrt{4V_{p}^2-\gamma_x^2},
\label{caleq1}
\end{eqnarray}
where $\gamma_x$ is the exciton linewidth and $V_{p}$ is the exciton-photon coupling strength. In typical GaAs-based structures, the coupling strength is of the order 8 meV, and exciton line widths in our system are around 8 meV as well. This means including the line widths in the calculation is important, as they represent a 20 to 25\% correction in Equation~(\ref{caleq1}). When the line widths are comparable to the splitting of the upper and polariton lines, however, the two-level model breaks down, since it is based on a perturbation-theory approach that assumes $\gamma \ll V_p$. 

We therefore use a more thorough approach to this problem based on computing the complex susceptiblity directly. This method starts with the transfer-matrix simulation code which successfully models the reflectivity spectra of our microcavity samples. It is possible to extract exciton fractions of the polaritons from this simulation, as discussed below.

\begin{figure}
\includegraphics[width=0.85\textwidth]{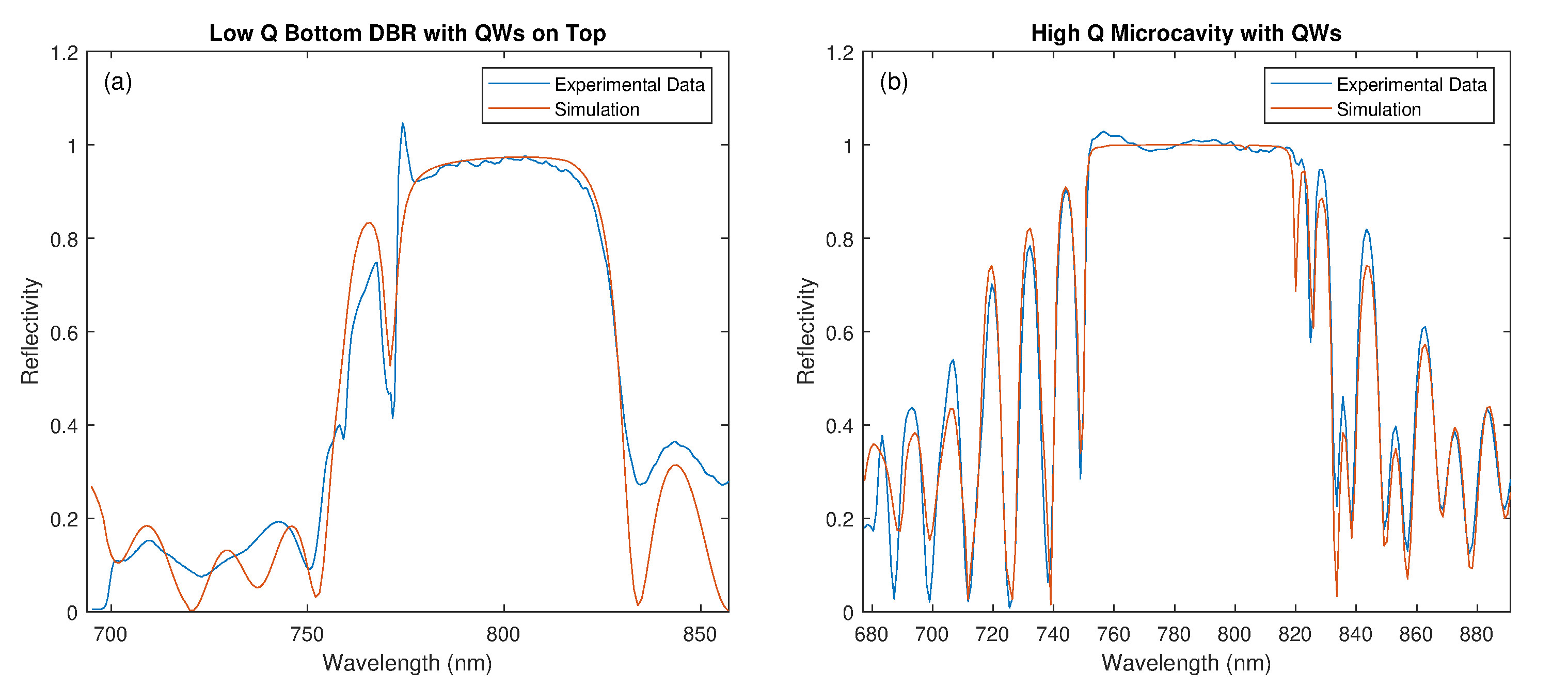}
\caption{(a) Comparison of our transfer matrix simulation to measured power reflectivity of a 20-period bottom DBR with quantum wells on top (Sample 9-3-14.1). This data allows us to get starting values for the exciton energy, line width, and the corresponding amplitude inside our electric susceptibility function for the excitons, given by Equation (\ref{calA2}). (b) Comparison of our transfer matrix simulation to measured power reflectivity off a high-$Q$ microcavity sample with quantum wells inside the cavity (Sample 4-6-15.1). This data allows us to fine tune the thicknesses and index values of the DBR layers. }
\label{refl}
\end{figure} 

In an ideal world, we could simply look up measured index values for our materials, and our transfer matrix simulation would match our experimental data. Unfortunately, there is not a lot of index data available for AlGaAs alloys at 4 K, and even if there were, the exact values can vary based on several factors, in particular the absorption coefficient and disorder due to impurities, which can change between growths, and can differ for different molecular-beam epitaxy machines. We start with the most accurate library we can find \cite{library}, and then using our measured reflectivity data, we uniformly scale the index functions and thicknesses of our DBR materials until the simulation and experimental reflectivity curves are in good agreement, as shown in Figure~\ref{refl}. 

The thickness and index scale factors are not entirely independent variables, as they are identical in an optical path length calculation. However, a critical factor in transfer matrix calculations is the ratio of the index values of the DBR materials, which determines the bandwidth of the stop band. If this ratio is maintained, scaling the thicknesses of all the index functions of the materials is an interchangeable action, and does not allow a unique fit. However, at larger angles of incidence this interchangeability breaks down, as the indices of the materials set the curvature of the Bragg modes. Comparing to angle-dependent data therefore allows a unique fit. We have found that doing a simultaneous fit of the reflectivity curves at 0, 15, and 30 degrees is sufficient to getting a unique fit of our DBR parameters. 

We also have found that using nominal book values \cite{library} for the imaginary component of the index never leads to accurate fits of our reflectivity data. The stop band we measure is always broader and flatter than the one we would simulate using such nominal values. We typically have to reduce this imaginary component by a factor of about ten in all DBR materials to get simulated stop bands resembling what we measure. It is worth noting that we use this same simulation for DBRs grown via PECVD, and do not need to reduce this imaginary component. This is likely a testament to the ultra-high purity of the samples grown by the Pfeiffer group at Princeton using MBE, which are almost certainly more defect-free than the samples used to measure the nominal book values. 

Once the DBR materials' thicknesses and index functions are scaled properly, we turn our attention towards the index function of the quantum wells inside the microcavity. Although there are many sophisticated models available in the literature, we chose to use a simple textbook model, as this simple model lends itself well to quick curve fitting. Following Ref.~\onlinecite{snokebook2}, Section 7.1, we model the contribution of the excitons to the complex electric susceptibility as a function of photon energy $E$ as  
\begin{eqnarray}
\chi(E) = \frac{A(E_x^2-E^2+i\gamma_x E)}{(E_x^2-E^2)^2+\gamma_x^2 E^2}
\label{calA2}
\end{eqnarray}
where $E_x$ is the exciton energy, $\gamma_x$ is the line width, and $A$ is related to the exciton-photon coupling strength $V_p$ (see Section 7.5.3 of Ref.~\onlinecite{snokebook2}). In principle, the exciton energy is angle-dependent, however in practice the excitons are so massive as to have an essentially flat dispersion in the energy range of interest. The dielectric function created by this susceptibility is then added to the dielectric function of bulk GaAs below the band gap, resulting in a final complex index for the quantum wells. Using a control sample consisting of bottom DBR with quantum wells on top, and no top DBR, we are able to get initial values for these three parameters. This data and the fit of the transfer-matrix model with the complex index of refraction are shown in Figure~\ref{refl}(a). 
 
\begin{figure}
\includegraphics[width=0.95\textwidth]{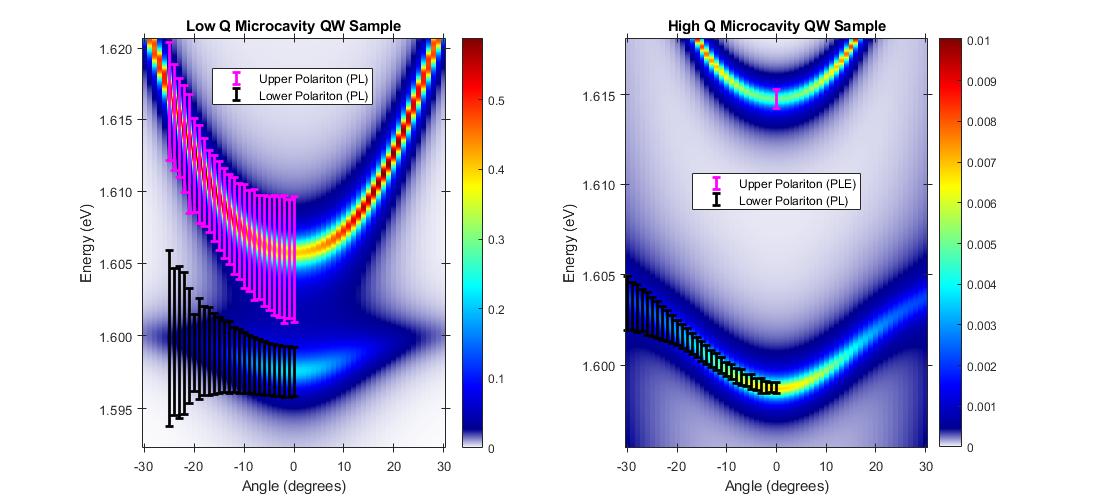}
\caption{Detailed comparison of the prediction of the model to experimental results. The background images give the reflectivity simulation results for $1-r$, using a narrow color scale to make the polariton dips more readily apparent. (a) Data (shown by error bars) and theory for a low-$Q$ microcavity sample (10-5-10.1), which has 20 periods in the top and bottom DBR, and only three quantum wells. The black and pink error bars correspond to the full width at half maximum for the PL peaks for the lower and upper polaritons, respectively, for non-resonant pumping ($\sim$720 nm, low-power CW laser). (b) Data (shown by error bars) and theory for a high-$Q$ microcavity sample (4-6-15.1), with 12 quantum wells. The lower polariton PL is visible with non-resonant pumping, but the upper polariton PL is unmeasurable due to down-conversion into lower polaritons via phonon emission, as discussed in Appendix A. Instead, the pink error bar corresponds to PLE measurement discussed in the text.}
\label{refl2}
\end{figure}

The three parameters in Equation (\ref{calA2}) only affect the quantum wells, which make up a small percentage of the total microcavity sample. So, in high-$Q$ samples, their value has virtually no effect on the broad characteristics of the curve shown in Figure~\ref{refl}(b). Additionally, we introduce one other parameter to scale the thickness of only the cavity layers, because this can vary across the wafers. This parameter also has almost no effect on the majority of the in curve in Figure~\ref{refl}(b). However, if we zoom in on that curve, there are two very narrow and shallow dips in the stop band corresponding to the polariton branches. Figure~\ref{refl2}(b) is a plot of $1-r$, where $r$ is the power reflectivity, at many angles, using a tight color scale to make these dips more visible. The four parameters of our model for the exciton susceptibility are almost entirely in control of the polariton modes while having almost no effect on the broad reflectivity curve. So long as the Bragg modes of the cavity do not begin intersecting the polariton branches, we have two nearly completely uncoupled fitting problems: polariton fitting inside the stopband and reflectivity fitting over the broader wavelength range.

With the broader reflectivity fitting already done using the DBR materials' parameters, we move on to tuning our four parameters for the exciton-polariton effect, namely the three parameters in Equation~(\ref{calA2}) and the total thickness of the cavity, to match our simulation of a low-$Q$ microcavity sample in Figure~\ref{refl2}(a). Because of the low $Q$-factor in this sample, we are able to experimentally extract the entire lower and upper polariton curves through both non-resonant pumping PL measurements and reflectivity measurements. This means in the low-$Q$ sample we have a larger data set to fit than in the high Q samples. Given this successful fit, we then apply it to the high-$Q$ sample, as shown in Figure~\ref{refl2}(b). The four parameters have to be tuned to match experimental data at each location on the sample, as the sample has varying thickness. We find the fitted line width of the exciton does not change much from location to location, and so effectively we have three parameters to tune to match the simulation to our experimental data of the two polariton branches, namely the exciton energy, the exciton-photon coupling strength $A$, and the thickness of the cavity (which corresponds to cavity mode's energy). 

Once all this fitting is done, and we have a simulation which agrees well with all of the experimental data, we can extract the lower polariton's exciton fraction from our simulation.  In the simple two-level model of Equation~\ref{polfrac}, one has simply 
\begin{eqnarray}
|\alpha|^2 = \frac{\partial E_{lp}}{\partial E_x}.
\label{hopdef}
\end{eqnarray}
We can therefore derive the value of the exciton fraction $|\alpha|^2$ (also often called $|X|^2$ in the literature) in the lower polariton state from our full complex susceptibility model by
slightly varying the exciton energy in Equation~\ref{calA2}, extracting the energy of our polariton dips, and finding the shift of the lower polariton as exciton energy varies.  This value is the crucial number for extracting the bare exciton-exciton interaction that gives a shift of the exciton energy. 

\begin{figure}
\includegraphics[width=0.95\textwidth]{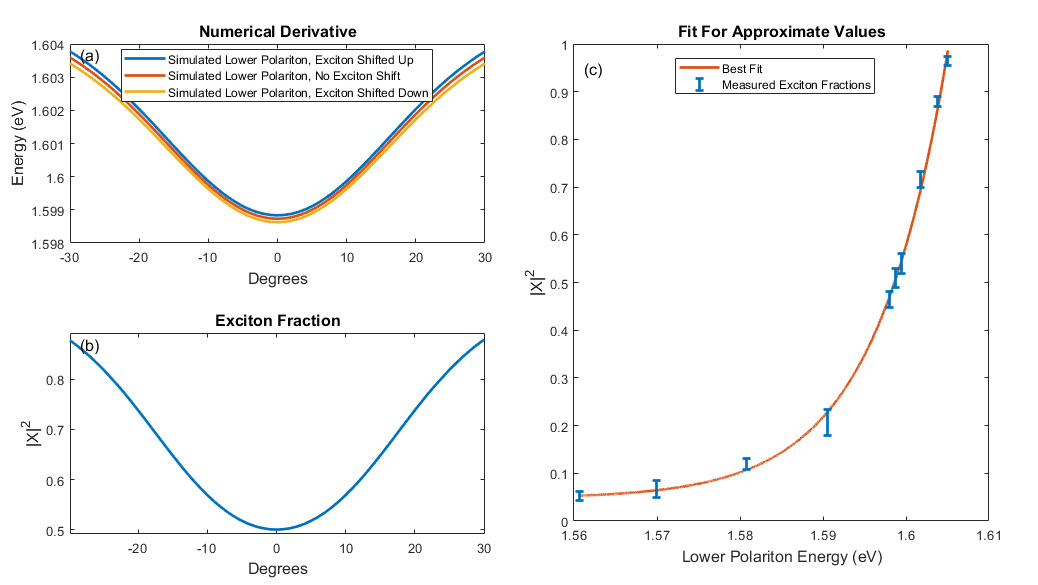}
\caption{(a) Red curve: results of our simulation after the polariton parameters have been fit to the experimental data for Sample 4-6-15.1. Blue and yellow curves: the same model with the exciton energy $E_x$ shifted slightly up and down. (b) The exciton fraction as defined in Equation (\ref{hopdef}) found by subtracting the curves of (a) and dividing by the difference in exciton energies. The lower polaritons are more excitonic at larger angles, as expected. (c) Blue data points: the results for the excitonic fraction found by from applying this procedure to several locations on the sample. In general the fits are more constrained near resonant detunings, hence the tighter error bars near 50\% exciton fraction. Red curve: The best fit of Equation~(\ref{calA4}) to these data. The fit parameters are given in Table ~\ref{fit_table}. }
\label{deriv}
\end{figure} 

In Figure \ref{deriv}(a) we can see how this small shift in exciton energy changes the lower polariton branch in the simulation. Then, by subtracting the two shifted curves to calculate the numerical derivative, we arrive at the exciton fraction of the lower polaritons plotted in Figure \ref{deriv}(b). As expected, this value is bounded by 0 and 1, approaching 1 at higher angles where the lower polariton becomes nearly purely excitonic. The results for the exciton fraction of the lower polariton line found using this procedure for the GaAs-based microcavity samples produced by the Pfeiffer group are shown in Figure~\ref{deriv}(c). 

It should be noted that this characterization procedure does not require data from multiple locations; it works at a single point on the sample. This is quite beneficial for certain applications and samples. However, the PLE measurement is cumbersome, and not all labs have access to the necessary tunable laser. We therefore fit our data to a simple curve given by
\begin{eqnarray}
|\alpha|^2 = \alpha_o + \alpha_s e^{\left({\frac{E_{lp} - E_r}{E_s}}\right)},
\label{calA4}
\end{eqnarray}
where $\alpha_o$, $\alpha_s$,$E_r$, and $E_s$ are fit parameters. The best fit with this function is given by the red line in Figure~\ref{deriv}(c). This figure shows the data for Sample 4-6-15.1, which was used in Refs.~\onlinecite{ostroTF} and \onlinecite{naturephys}. Additionally, we performed the same procedure on sample P8-10-17.1, which was used in Ref.~\onlinecite{myers-pillar}. The parameter values for the best fits are given in Table~\ref{fit_table}. \\

\begin{table}
\begin{tabular}{|c|c|c|c|c|}
\hline
\hspace{.2cm} Sample \hspace{.2cm}  & \hspace{.3cm}  $\alpha_o$ \hspace{.3cm}   & \hspace{.3cm} $\alpha_s$   \hspace{.3cm}  & \hspace{0cm}  $E_r$ (eV) \hspace{0cm} &  \hspace{0cm} $E_s$ (meV)  \hspace{0cm} \\ \hline
4-6-15.1 & 0.0468 & 3.3373 &  1.6162 & 8.83 \\
P8-10-17.1  & 0.1758 &  3.3832 & 1.6221 & 11.4 \\
\hline
\end{tabular}
\caption{Fitted parameters for equation ~\ref{calA4}}
\label{fit_table}
\end{table}

{\bf Calibration of previous work}. In the past, less accurate characterizations were utilized.  We therefore compare some of our prior characterizations with the results given here. 

In Reference~\onlinecite{naturephys}, Figure 2, the authors stated that the detuning was 7.7 meV, however using our present calibration, polaritons at that energy have a predicted detuning of 17.8 meV (exciton fraction of 85\%).  Additionally, in Figure 4 of that paper, an approximate formula for the exciton fraction was utilized, in which very photonic polaritons (detuning of $-22.5$ meV) were approximated as being bare cavity photons, giving the equation $|\alpha|^2 \simeq 1 - m_c/m_{lp}$, where $m_c$ and $m_{lp}$ are the measured cavity mass and lower polariton mass, respectively.  However, at such a detuning, the exciton fraction is 11\%, which means the true cavity mass should be about 89\% of the cavity mass which was used. This correction shifts all data points in the figure to the right in a non-uniform way. A data point at $x = 0.2$ would shift to $x = 0.3$, and a data point at $x = 0.8$ would shift to $x = 0.826$. 

In Reference \onlinecite{myers-pillar}, the data presented in Figure 2, according to our current calibration, is at a detuning of 11.7 meV; contrasted with the 8 meV based on the previous calibration method. The overall conclusions of that paper are not significantly changed.

In Reference \onlinecite{ostroTF},  Figure 1, the authors state that the detuning of their polaritons is 2~meV; based on our current calibration, we find a detuning of 3.4 meV.  In Figure 2(a) of that paper, the stated detuning was $-18$ meV, close to our current value of $-18.2$ meV.  In Figure 2(c), the stated detuning was 0 meV, while our current calibration gives a value of $0.2$ meV.  In Figure 2(e), the stated detuning was 20 meV, while our current calibration gives 17.9 meV. The overall conclusions of that paper are therefore not significantly changed.

\section{Updated quantum kinetic model of polariton equilibration}
\label{apphart}

\begin{figure}[b]
\includegraphics[width=0.75\textwidth]{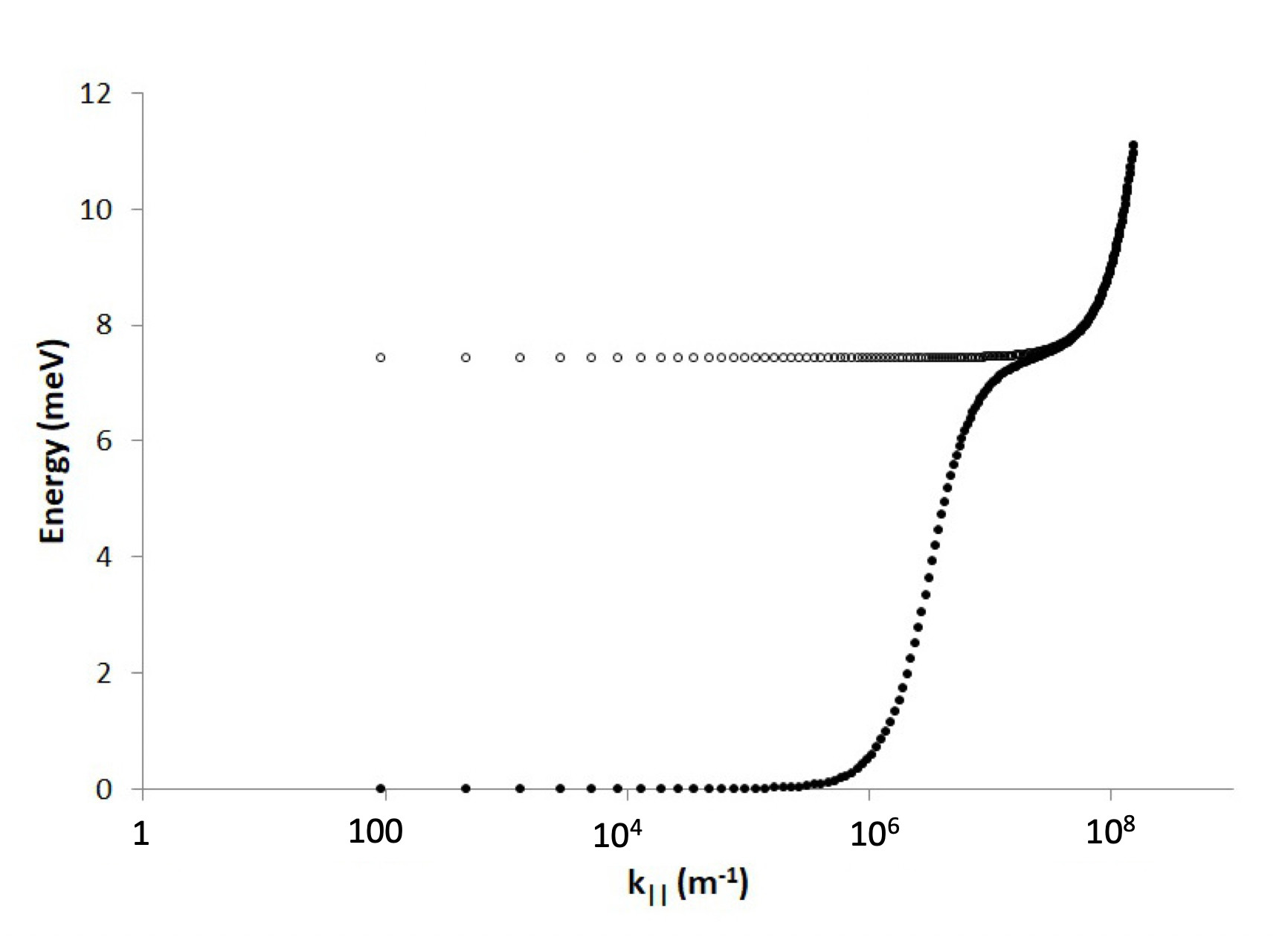}
\caption{The dispersion curves for the polaritons (filled circles) and bare excitons (open circles).}
\label{Esteps}
\end{figure}

The code of Ref.~\onlinecite{hartwell} has been updated to account for both dark excitons, which do not couple to light due to selection rules, but can convert to and from bright excitons via transverse phonon absorption and emission, and scatter with excitons and polaritons. In this model,  the lower polariton band which evolves continuously into the bright exciton state; there is no formal separation of the polariton population and the bright exciton population. The two bands (dark excitons and bright-exciton/polariton) use the same mesh in $k$-space and each uses its respective dispersion curve.  The two dispersion curves overlap for large $k$ as illustrated in Figure~\ref{Esteps}.

The general process of the simulation is the same as the polariton-only simulation described in Ref.~\onlinecite{hartwell}.  Adding the new population requires a second Boltzmann equation, so that we have
\begin{eqnarray}
\frac{\partial n_{\vec{k}}^{lp}}{\partial t} &=& P_{\vec{k}}(t) - \frac{n_{\vec{k}}^{lp} }{\tau_{\vec{k}} } + \sum_{\vec{k'}} W_{\vec{k}'\rightarrow\vec{k}}^{(i)}(t) - \sum_{\vec{k}'}W_{\vec{k}\rightarrow\vec{k}'}^{(i)}(t),\nonumber\\
\label{eq1}
\end{eqnarray}
\begin{eqnarray}
\frac{\partial n_{\vec{k}}^{Dx}}{\partial t} &=& -\frac{n_{\vec{k}}^{Dx} }{\tau_{\vec{k}} } + \sum_{\vec{k'}} W_{\vec{k}'\rightarrow\vec{k}}^{(i)}(t) - \sum_{\vec{k}'}W_{\vec{k}\rightarrow\vec{k}'}^{(i)}(t),\nonumber\\
\label{eq2}
\end{eqnarray}
where $n_{\vec{k}}$ is the occupation number, 	$\tau$ is the characteristic lifetime, $P_{\vec{k}}$ is the pumping term, and 
the label $lp$ refers to the lower polariton branch,
and the label $Dx$ refers to the dark exciton population. The values of $\vec{k}$ and $\vec{k'}$ range over each value of the mesh.  
The $W^{(i)\prime}s$ are interaction terms for particle-particle interactions within each population, the particle-longitudinal acoustical phonons interactions for a population, the particle-transverse acoustical phonon interactions for a population, and the newly added particle-transverse acoustical phonon interactions between populations. 
In the last case, $\vec{k}$ belongs to one population and $\vec{k'}$ belongs to the other, where both values simultaneously meet the energy requirement stated above.

The time step of this model is determined as in Ref.~\onlinecite{hartwell} and is based on moving a small fraction of the polariton population each iteration.  That time step is then directly multiplied by Equation \eqref{eq2} to determine the change in occupation number of the dark excitons for the iteration being performed.

The model had three free parameters which could be varied to fit the experimental results: the lattice temperature (which was constrained to be no greater than the experimentally measured effective polariton temperature), the bare exciton-exciton scattering parameter $g_{ex}$, and the background free electron density.  The cavity lifetimes had one of two values: 1 ps, for the ``short lifetime'' experiments (e.g., Refs.~\onlinecite{science} and \onlinecite{hartwell}), and 100 ps, for the ``long lifetime'' experiments, (e.g., Refs. \onlinecite{naturephys} and \onlinecite{equilprl}).  The experimental constraints results that needed to be modeled with consistent parameters included the following: a) For the short-lifetime experiments, both the occurrence of BEC at high density, and the nearly unchanging distribution of $n_{\vec{k}}$ at low density had to be reproduced. As noted in the main text, this was only possible for a high value of $g_{ex}$; as discussed in the main text, taking the nominal theoretical value for $g$ badly misses the experimental result; b) for the long-lifetime experiments, the approximately linear growth of the polariton density as pump power was increased, well below the BEC threshold, had to reproduced; and c) the interaction parameter $g_{ex}$ had to be the same for both the short-lifetime and long-lifetime data, since the structures used in these experiments had the same quantum well structure, and differed only in the cavity lifetime.  Since collisional processes scale as the square of density, the linear growth of the polariton density seen experimentally meant that conversion processes of excitons into polaritons used in our simulations had to be dominated by phonon and free-electron scattering at low density.  We found that these could only be reproduced by including a background free electron gas with a density of around $10^8$ cm$^{-2}$.

\section{Many-body screening effects}
\label{appscreen}

The hard-core repulsive interactions of the polaritons lead to a renormalized scattering interaction strength similar to how the repulsive interactions of electrons lead to screening. In the case of bosons with hard-core interactions, the term ``screening'' may not be appropriate, since that term is often associated with a long-range Coulomb interaction. However, 
the standard Lindhard screening model can be directly adapted to a Bose gas.  As discussed in the main text, the effect is fundamentally due to spatial anti-correlation of particles with a repulsive interaction, which spend less time near each other, and therefore give an effectively lower density felt by any single particle.  The diagrams for bosons are the same as for fermions; the only difference is that the interaction matrix element is changed from the Coulomb electron-electron interaction to a hard-core repulsion. 
\begin{figure}
\includegraphics[width = .2\linewidth]{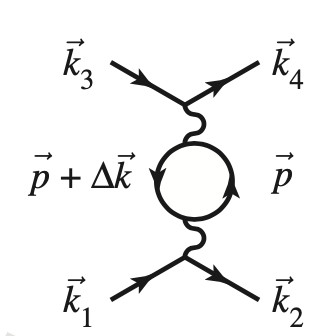}
\caption{Diagram for second-order renormalization of the particle-particle interaction.} 
\label{screen1}
\end{figure}

Following the procedure of Ref.~\onlinecite{fetter},  Sections 9, 30 and 35, the effective interaction vertex of bosons is (see Equation (9.45) and Problem 5.8 on page 196)
\begin{eqnarray}
\tilde{g}' \simeq \tilde{g} \frac{1- \tilde{g}\Pi^0}{1-2\tilde{g}\Pi^0},
\label{fetter1}
\end{eqnarray}
where $\Pi^0$ is the polarization bubble taking into account many-body screening effects, shown in Figure \ref{screen1}.
Using the Matsubara finite-temperature approach for bosons, discussed in Section 8.15 of Ref.~\onlinecite{snokebook2}, we write for this bubble \begin{eqnarray}
\Pi_{\Delta\vec{k}} &=&  -\sum_{\vec{p}} \frac{1}{\beta}\sum_{m=-\infty}^{\infty} \frac{1}{(i\hbar\omega_m  - E_{\vec{p}} + \mu)}
\frac{1}{(i\hbar\omega_m +i\hbar\omega_n - E_{\vec{p}+\Delta\vec{k}} + \mu)},
\label{matsupol}
\end{eqnarray}
where $m$ and $n$ are even integers.
Resolving the sum over $m$ using the standard boson sum rule gives
\begin{eqnarray}
\Pi_{\vec{k},n} &=&  \sum_{\vec{p}}\left( \frac{N_B(E_{\vec{p}}-\mu)}{i\hbar\omega_n +E_{\vec{p}}-E_{\vec{p}+\Delta\vec{k}} }
+\frac{N_B(E_{\vec{p}+\Delta\vec{k}}-\mu-i\hbar\omega_n)}{E_{\vec{p}+\Delta\vec{k}}  -i\hbar\omega_n -E_{\vec{p}}}
 \right) \nonumber\\
 &=&  \sum_{\vec{p}}\left( \frac{N_B(E_{\vec{p}}-\mu)}{E_{\vec{p}}-E_{\vec{p}+\Delta\vec{k}} + i\hbar\omega_n  }
-\frac{N_B(E_{\vec{p}+\Delta\vec{k}}-\mu)}{E_{\vec{p}} - E_{\vec{p}+\Delta\vec{k}}  +i\hbar\omega_n }
 \right), 
\end{eqnarray}
where $N_B$ is the Bose-Einstein distribution of the particles for chemical potential $\mu$ and temperature $T$, and we have used the rule that $N_B(E_{\vec{k}} + i\hbar\omega_n) = N_B(E_{\vec{k}})$ for even $n$. Using analytic continuation $i\omega_n \rightarrow \omega + i\epsilon$, and writing $\Delta E = \hbar\omega$, we then have
\begin{eqnarray}
\Pi_{\Delta\vec{k}}  = \sum_{\vec{p}} \frac{N_{\vec{p}}-N_{\vec{p}+\Delta\vec{k} } }{\Delta E + E_{\vec{p}} - E_{\vec{p}+\Delta\vec{k}} +i\epsilon }.
\end{eqnarray}
This is the Lindhard formula for bosons. In the long wavelength, low frequency limit, this becomes
\begin{eqnarray}
\Pi_{\Delta k}  &=& \sum_{\vec{p}} \frac{\partial N_B(E_p)}{\partial E_p} \nonumber\\
&=& \frac{A m}{\pi \hbar^2}\int_0^\infty dE \ \frac{\partial N_B}{\partial E} \nonumber\\ 
&=&  \frac{A m}{\pi \hbar^2} (N_B(\infty)-N_B(0)) \nonumber\\ 
&=& -\frac{A m}{\pi \hbar^2}\frac{1}{e^{-\mu/k_BT}-1}.
\end{eqnarray}
where we have used the two-dimensional density of states for a gas with two-fold degeneracy and area $A$.

In the condensate regime, this becomes large and negative. Then (\ref{fetter1}) becomes $\tilde{g}' \rightarrow \tilde{g}/2$. This is consistent with the approximate calculation of Section 11.2 of Ref.~\onlinecite{snokebook2} which found a factor of 2 reduction of the interaction energy in going from a thermal gas to a condensate, due to bosonic exchange.





\begin{thebibliography}{99}

\bibitem{littlewood} P.B. Littlewood and A. Edelman, in {\em Universal Themes of Bose-Einstein Condensation}, (Cambridge University Press, 2017). 

\bibitem{keeling} D.W. Snoke and J. Keeling, Physics Today {\bf 70}, 54  (October 2017). 

\bibitem{Delteil2019} A. Delteil, Th. Fink, A. Schade, S. H\"ofling, C. Schneider, and
A. Imamoglu, 
Nat. Mater. {\bf 18}, 219 (2019).

\bibitem{Matutano2019} G. Munoz-Matutano, A. Wood, M. Johnson, X. Vidal Asensio,
B. Baragiola, A. Reinhard, A. Lemaitre, J. Bloch, A. Amo,
B. Besga, M. Richard, and T. Volz, 
Nat. Mater. {\bf 18}, 213 (2019).

\bibitem{baum} T. Gao, P. S. Eldridge, T. C. H. Liew, S. I. Tsintzos, G. Stavrinidis, G. Deligeorgis, Z. Hatzopoulos, and P. G. Savvidis, Phys. Rev. B {\bf 85}, 235102 (2012).


\bibitem{ballarini} D. Ballarini, M. De Giorgi, E. Cancellieri, R. Houdre, E. Giacobino, R. Cingolani, A. Bramati, G. Gigli and D. Sanvitto,  Nature Comm. {\bf 4}, 1778 (2013).

\bibitem{recentgate} A.V. Zasedatelev, A.V. Baranikov, D. Urbonas, F. Scafirimuto, U. Scherf, T. Stoeferle, R.F. Mahrt and P.G. Lagoudakis, Nature Photonics {\bf 13}, 378 (2019).

\bibitem{snokebook2} D.W. Snoke, {\em Solid State Physics: Essential Concepts}, 2nd edition (Cambridge University Press, 2019).

\bibitem{tassone} F. Tassone and Y. Yamamoto, Phys. Rev. B {\bf 59}, 10830 (1999). 

\bibitem{ciuti} C. Ciuti, V. Savona, C. Piermarocchi, A. Quattropani, and P. Schwendimann,  
Phys. Rev. B {\bf 58}, 7926 (1998).

\bibitem{ciuti2} A. Verger, C. Ciuti, and I. Carusotto, Phys. Rev. B {\bf 73}, 193306 (2006).

\bibitem{wouters} 
M. Wouters, Phys. Rev.  B {\bf 76}, 045319 (2007).

\bibitem{combescot} M. Combescot and O. Betbeder-Matibet,
Solid State Comm. {\bf 134}, 11 (2005).


\bibitem{yamamoto} H. Deng, G. Weihs, C. Santori, J. Bloch, and Y. Yamamoto,  Science {\bf 298}, 199 (2002).

\bibitem{science} R. Balili, V. Hartwell, D.W. Snoke,  L. Pfeiffer and K. West,  Science {\bf 316}, 1007 (2007).

\bibitem{bloch} M. Maragkou, A. J. D. Grundy, E. Wertz, A. Lemaitre, I. Sagnes, P. Senellart, J. Bloch, and P. G. Lagoudakis, Phys. Rev. B {\bf 81}, 081307(R) (2010).

\bibitem{vlad} 
M. Vladimirova, S. Cronenberger, D. Scalbert, M. Nawrocki, A. V. Kavokin, A. Miard, A Lemaitre, and J. Bloch
Phys. Rev. B {\bf 79}, 115325 (2009).

\bibitem{ferrier} L. Ferrier, E. Wertz, R. Johne, D.D. Solnyshkov, P. Senellart, I. Sagnes, A. Lemaitre, G. Malpuech, and J. Bloch,
 Phys. Rev. Lett. {\bf 106}, 126401 (2011).

\bibitem{walker} P. M. Walker, L. Tinkler, B. Royall, D. V. Skryabin, I. Farrer, D. A. Ritchie, M. S. Skolnick, and D. N. Krizhanovskii, Phys. Rev. Lett. {\bf 119}, 097403 (2017).

\bibitem{devres} N. Takemura, S. Trebaol, M. Wouters, M. T. Portella-Oberli, and B. Deveaud, Phys. Rev. B {\bf 90}, 195307 (2014).

\bibitem{devres2} N. Takemura, M.D. Anderson, S. Biswas, M. Navadeh-Toupchi, D.Y. Oberli, M.T. Portella-Oberli, and B. Deveaud, Phys. Rev. B {\bf 94}, 185301 (2016).



\bibitem{myers-pillar} D. Myers, J. Wuenschell, D. Snoke, L. Pfeiffer, and K. West,  Applied Phys. Lett. {\bf 110}, 211104 (2017).

\bibitem{ostroTF} E. Estrecho, T. Gao, N. Bobrovska, D. Comber-Todd, M. D. Fraser, M. Steger, K. West, L. N. Pfeiffer, J. Levinsen, M. M. Parish, T. C. H. Liew, M. Matuszewski, D. W. Snoke, A.G. Truscott, and E. A. Ostrovskaya, Phys. Rev. B  {\bf 100}, 035306 (2019).

\bibitem{ostrobogo} M. Pieczarka, E. Estrecho, M. Boozarjmehr, O. Bleu, M. Steger, K. West, L. N. Pfeiffer, D. W. Snoke, J. Levinsen, M. Parish, A. G. Truscott and E. A. Ostrovskaya
Nature Comm. {\bf 11}, 429 (2020).

\bibitem{exdiff} D.M. Myers, S. Mukherjee, J. Beaumariage, D. W. Snoke, M. Steger, L. N. Pfeiffer and K. West, 
Phys. Rev. B {\bf 98}, 235302 (2018).


\bibitem{naturephys} Y. Sun, Y. Yoon, M. Steger, G.-Q. Liu, L.N. Pfeiffer, K. West, D.W. Snoke, and K.A. Nelson,  Nature Physics {\bf 13}, 870 (2017).

\bibitem{baum-shift} G. Tosi, G. Christmann, N. G. Berloff, P. Tsotsis, T. Gao, Z. Hatzopoulos, P. G. Savvidis,
and J. J. Baumberg, Nature Phys. {\bf 8}, 190 (2012).

\bibitem{piec} M. Pieczarka, M. Boozarjmehr, E. Estrecho, Y. Yoon, M. Steger, K. West, L. N. Pfeiffer, K. A. Nelson, D. W. Snoke, A. G. Truscott, E. A. Ostrovskaya,  Phys. Rev. B {\bf 100}, 085301 (2019). 

\bibitem{QWexdiff} H. Hillmer, A. Forchel, R. Sauer, and C.W. Tu, Phys. Rev. B {\bf 42}, R3220 (1990).

\bibitem{mass} K.-S. Lee and E.-H. Lee, ETRI Journal {\bf 17}, 13 (1996).

\bibitem{scurve} E. Wertz, et al., Applied Phys. Lett. 
{\bf 95}, 051108 (2009).


\bibitem{brichkin}  A. S. Brichkin, S. I. Novikov, A. V. Larionov, V. D. Kulakovskii, M. M. Glazov, C. Schneider, S. H\"ofling, M. Kamp, and A. Forchel,
Phys. Rev. B {\bf 84}, 195301 (2011).


 \bibitem{ring} S. Mukherjee, D.M. Myers, R.G. Lena, B. Ozden, J. Beaumariage, Z. Sun, M. Steger, L.N. Pfeiffer, K. West, A.J. Daley, and D. W. Snoke,  Phys. Rev. B {\bf 100}, 245304 (2019).


\bibitem{hartwell} V. E. Hartwell and D.W. Snoke,  Phys. Rev. B {\bf 82}, 075307 (2010).

\bibitem{equilprl} Y. Sun, P. Wen, Y. Yoon, G.-Q. Liu,  M. Steger, L.N. Pfeiffer, K. West, D.W. Snoke, and K.A. Nelson,  Phys. Rev. Lett. {\bf 118}, 016602 (2017). 


\bibitem{block1} A. Verger, C. Ciuti, C. and I. Carusotto, 
Phys. Rev. B {\bf 73}, 193306 (2006).

\bibitem{block2}
S. Ferretti and D. Gerace. 
Phys. Rev. B {\bf 85}, 033303 (2012).

\bibitem{boser} H. Cao, S. Pau, J. M. Jacobson, G. Bj\"ork, Y. Yamamoto, and A. Imamoglu, Phys. Rev. A {\bf 55}, 4632 (1997).

\bibitem{jap} B. Nelsen, R. Balili, D.W. Snoke, L. Pfeiffer, and K. West, 
J. Applied Phys. {\bf 105}, 122414 (2009).

\bibitem{bloch2017}  S. R. K. Rodriguez, A. Amo, I. Sagnes, L. Le Gratiet, E. Galopin, A. Lemaitre and J. Bloch,  
Nat. Commun. {\bf 7}, 11887 (2016).

\bibitem{zimm}  R. Zimmermann, Phys. stat. sol. (b) {\bf 243}, 2358 (2006).

\bibitem{laikht} 
B. Laikhtman and R. Rapaport,  Europhys. Lett. {\bf 87}, 27010 (2009).

\bibitem{fesh} O. Bleu, J. Levinsen, M.M. Parish, ``Interplay between polarization and quantum correlations of confined polaritons,'' arXiv:2104.13541.

\bibitem{keeling-private} J. Keeling, private communication. 

\bibitem{shouvik} S. Mukherjee, D. M. Myers, R. G. Lena, B. Ozden, J. Beaumariage, Z. Sun, M. Steger, L. N. Pfeiffer, K. West, A. J. Daley, and D. W. Snoke,
Phys. Rev. B {\bf 100}, 245304 (2019) 

\bibitem{skolT2} A. P. D. Love et al., 
Phys. Rev. Lett. {\bf 101}, 067404 (2008).


\bibitem{prx} B. Nelsen, G-Q. Liu, M. Steger, D.W. Snoke, R. Balili, K. West and L.N. Pfeiffer,  Physical Review X {\bf 3}, 041015 (2013).

\bibitem{steger-prb} M. Steger, G.-Q. Liu, B. Nelsen, C. Gautham, D.W. Snoke, R. Balili, L.N. Pfeiffer and K. West,  Phys. Rev. B {\bf 88}, 235314 (2013). 

\bibitem{SR} S. Schmitt-Rink, D.S. Chemla, and D.A.B. Miller, Phys. Rev. B {\bf 32}, 6601 (1985).

\bibitem{exbohr} K. Litvinenko, D. Birkedal, V. G. Lyssenko, and J. M. Hvam, Phys. Rev. B {\bf 59}, 10255 (1999).

\bibitem{library} J.A. Gonzalez-Cuevas, J. Applied Phys. {\bf 102}, 014504 (2007). 


\bibitem{ash} A. Askitopoulos, L. Pickup, S. Alyatkin, A. Zasedatelev, K.G. Lagoudakis, W. Langbein, and P.G. Lagoudakis, ``Giant increase of temporal coherence in optically trapped
polariton condensate,'' arXiv:1911.08981.


\bibitem{fetter} A.L. Fetter and J.D. Walecka, {\em Quantum Theory of Many-Particle Systems}, (Dover, 2003).


\end{thebibliography}
\end{document}